\documentclass[aps,onecolumn,preprint,superscriptaddress,nofootinbib,floats]{revtex4}
\usepackage{amsmath,amssymb,color,mathrsfs, graphicx,verbatim,epsfig, bbm, wasysym}

\def\lsim{\mathrel{\rlap{\lower4pt\hbox{\hskip1pt$\sim$}}
    \raise1pt\hbox{$<$}}}
\def\gsim{\mathrel{\rlap{\lower4pt\hbox{\hskip1pt$\sim$}}
    \raise1pt\hbox{$>$}}}

\newcommand{\be}{\begin{eqnarray}}
\newcommand{\ee}{\end{eqnarray}}

\def\addresses#1#2{\hbox to \hsize{\@tablebox{#1}\hfil\@tablebox{#2}}}
\def\@tablebox#1{\vtop{\hsize=5in \begin{flushleft} #1 \end{flushleft}}}

\def\beq{\begin{equation}}
\def\eeq{\end{equation}}
\def\bit{\begin{itemize}}
\def\eit{\end{itemize}}
\def\beqarray{\begin{eqnarray}}
\def\eeqarray{\end{eqnarray}}
\def\ttbar{$t\overline{t}$}
\def\ljets{$l$+jets}

\def\mujets{$\mu$+jets}

\def\bjet{$b$-jet}
\def\DR{$\Delta R$}
\def\DeltaR{\Delta R}
\def\DRbmu{$\Delta R_{b\mu}$}
\def\met{$\displaystyle{\not}E_T$}
\def\PYTHIA{{\tt PYTHIA}}
\def\HERWIG{{\tt HERWIG}}
\def\MadGraph{{\tt MadGraph}}
\def\MadEvent{{\tt MadEvent}}
\def\FastJet{{\tt FastJet}}

\begin{document}

\baselineskip 0.6cm

\begin{titlepage}

\thispagestyle{empty}

\begin{flushright}
\end{flushright}

\begin{center}

\vskip 2cm

{\Large \bf Efficient Identification of Boosted Semileptonic \\ \bf Top Quarks at the LHC}
\vskip 1.0cm
{\large Keith Rehermann and Brock Tweedie}
\vskip 0.4cm
{\it Department of Physics and Astronomy, Johns Hopkins University,
           Baltimore, MD 21218} \\
\vskip 1.2cm

\end{center}

\noindent Top quarks produced in multi-TeV processes will have large Lorentz boosts, and their decay products will be highly collimated.  In semileptonic decay modes, this often leads to the merging of the $b$-jet and the hard lepton according to standard event reconstructions, which can complicate new physics searches.  Here we explore ways of efficiently recovering this signal in the muon channel at the LHC.  We perform a particle-level study of events with muons produced inside of boosted tops, as well as in generic QCD jets and from $W$-strahlung off of hard quarks.  We characterize the discriminating power of cuts previously explored in the literature, as well two new ones.  We find a particularly powerful isolation variable which can potentially reject light QCD jets with hard embedded muons at the $10^3$ level while retaining 80$\sim$90\% of the tops.  This can also be fruitfully combined with other cuts for $O(1)$ greater discrimination.  For $W$-strahlung, a simple $p_T$-scaled {\it maximum} \DR\ cut performs comparably to a highly idealized top-mass reconstruction, rejecting an $O(1)$ fraction of the background with percent-scale loss of signal.  Using these results, we suggest a set of well-motivated baseline cuts for any physics analysis involving semileptonic top quarks at TeV-scale momenta, using neither $b$-tagging nor missing energy as discriminators.  We demonstrate the utility of our cuts in searching for resonances in the \ttbar\ invariant mass spectrum.  For example, our results suggest that 100 fb$^{-1}$ of data from a 14 TeV LHC could be used to discover a warped KK gluon up to $4.5$ TeV or higher.

\end{titlepage}

\setcounter{page}{1}

\section{Introduction}

The LHC promises our first glimpse of top quarks produced at energies far above threshold.  There has been much speculation about top's role at these energies, given its large coupling to the sector that breaks electroweak symmetry.  For example, models which complete the electroweak sector with strong dynamics often contain a rich spectrum of heavy composite resonances with large branching fractions to top quarks.  Given the constraints on these models from flavor and electroweak precision tests (see e.g., \cite{Agashe:2008uz,Agashe:2003zs}), as well as from direct searches at the Tevatron \cite{Aaltonen:2009tx,Abazov:2008ny}, it is generally expected that the lowest-lying composite states must live in the multi-TeV mass range.  Top quarks produced in the decays of such massive particles will be so energetic and highly Lorentz-boosted that they effectively inhabit a new kinematic regime, where all of an individual top's decay products can become beamed into a localized region of the detector---a ``top-jet.''   Our ability to discover hints of compositeness within the electroweak sector, or to place significant constraints on it, may therefore depend on how reliably we can reconstruct these highly boosted tops.  More generally, final states with boosted tops will serve as a probe of any new multi-TeV-scale physics that can couple to top quarks.

Reconstructing tops at high boost comes with special challenges, since the $O(1)$ angles between decay products in the top rest frame to shrink to \DR\ $\sim 0.1$ in the lab.  Conventional searches will often group together two or more of these particles into a single jet, losing information on internal kinematics.  Further complicating matters, $b$-jets face degraded tagging efficiency due to the high density of nearly-collinear tracking hits in the inner detectors (see, e.g., \cite{Garcia:xxx}).  Missing transverse energy may also be difficult to utilize for detailed top kinematic reconstruction since it is nearly aligned with the lepton and the $b$-jet, and is therefore particularly sensitive to fluctuations in visible energy measurements.

In a previous publication \cite{Kaplan:2008ie}, we and our collaborators demonstrated that these difficulties can plausibly be overcome for the case of boosted tops that decay hadronically.  (For similar ideas, see also \cite{Brooijmans:2008zza,Thaler:2008ju,Almeida:2008yp,Ellis:2009su,Chekanov:2010vc}.)  The three jets produced in the top's decay merge according to traditional jet definitions, but still leave patterns in the calorimeter that can be discriminated from jets initiated by light quarks or gluons.  This calorimeter-based ``top-tag'' potentially outperfoms $b$-tagging for $p_T \sim$ TeV, achieving efficiencies similar to $b$-tagging at much lower $p_T$.  This opens up the possibility of performing a search for resonances in the \ttbar\ mass spectrum utilizing the all-hadronic decay channel, independent of whether $b$-tagging can be made to work effectively at TeV-scale momenta.  For recent studies of these techniques in full detector simulation at CMS, see \cite{Rappoccio:2009cr,CMS:2009a,CMS:2009b}.

Boosted tops that decay semileptonically should be easier to identify, as they feature hard leptons and missing energy.  Indeed, most work to date on searching for multi-TeV \ttbar\ resonances, such as \cite{Barger:2006hm,Agashe:2006hk,Lillie:2007yh,Baur:2007ck,Baur:2008uv}, has focused on the \ljets\ channel because it is generally considered to be an optimal compromise between signal rate and background discrimination.  However, at high top boost we inevitably face the question of how to handle leptons that technically live inside of jets.  This is essentially a new problem for multi-TeV machines, since earlier colliders such as the Tevatron would produce leptons in jets mainly through heavy flavor decays or by geometric accidents (or fakes).  Traditionally, non-isolated leptons are only considered for use in $b$-tagging, and not as independently reconstructable objects.

This issue has been dealt with in several ways in the theory literature.  The investigation of top resonances in \cite{Baur:2007ck,Baur:2008uv} used conservative lepton isolation criteria, but at significant cost of signal efficiency at the highest masses.  In \cite{Agashe:2006hk}, a lepton could either be isolated or have a high invariant mass when combined with its host jet.  In \cite{Thaler:2008ju}, a suite of simple kinematic cuts were developed to help identify leptons ``stuck'' inside of jets.

In this paper we investigate the discriminating power of many of the previously suggested variables, as well as two new, especially simple and powerful ones.  From this investigation, we propose a minimal set of cuts for discriminating semileptonic boosted tops from their two major physics backgrounds:  heavy flavor pair-production within jets and $W$-strahlung off of hard quarks.  We focus on the relatively clean case of muons.  We find that the heavy flavor background can be very efficiently removed with a novel tracking isolation cut, tallying tracker energy in a small isolation cone that shrinks with muon $p_T$.  This variable is sensitive to the nearby showering and decay products of high-$p_T$ bottom and charm quarks.  This becomes more powerful in combination with a cut on the muonic mass-drop variable $x_\mu$, or on $\Delta R_{b\mu}$, both of which where investigated in \cite{Thaler:2008ju}.  In principle, this combination allows a rejection of QCD jets with embedded leptons at the level of $10^3$, and QCD jets in general at the level of $10^4 \sim 10^5$, with only $O(10\%)$ loss of signal.  For $W$-strahlung, we find that a simple cut on the $p_T$-scaled \DR\ between the muon and the closest jet performs essentially as well as an idealized top invariant mass cut using a perfectly measured neutrino three-vector.  An $O(1)$ rejection of $W$-strahlung events can be achieved with a loss of a few percent of the signal.  Throughout, we avoid utilizing any parametrization of $b$-tagging, and conservatively avoid using \met\ for the construction of discriminator variables.

As a test case for our methods, we investigate signals and backgrounds for chiral \ttbar\ resonances in the \mujets\ channel.  With our cuts, in combination with modest additional cuts on the hadronic side, the heavy flavor background is brought down to a level roughly $1 \sim 2$ orders of magnitude below the irreducible \ttbar\ background.  The $W$-strahlung background remains important, becoming dominant above about $2.5$ TeV invariant mass if no additional discrimination methods are used.  Using the full hadronic top-tag of \cite{Kaplan:2008ie}, $W$-strahlung can be made completely subdominant, but at an additional $O(1 \sim 10)$ cost in signal efficiency.  We also demonstrate that top polarization information, encoded in the relative muon momentum, can still be utilized with our semileptonic cuts.

In section \ref{sec:heavyflavor}, we discuss the physics motivation and estimate the performance of discriminator variables useful for eliminating jets containing heavy flavor.  In section \ref{sec:Wstrahlung}, we discuss $W$-strahlung and how to efficiently discriminate against it.  Section \ref{sec:resonances} presents the backgrounds in the \ttbar\ invariant mass spectrum using our methods, and estimates discovery reach for some simple models using a nominal set of cuts.  We present conclusions in section \ref{sec:conclusions}.

\section{Heavy Flavor}

\label{sec:heavyflavor}

\subsection{Leptons inside of jets}

\label{subsec:lepsinjets}

The lepton and the \bjet\ generated in the semileptonic decay of a boosted top will often overlap according to standard event reconstructions.  More specifically, for left/right chirality top quarks produced at $p_T \simeq 1$ TeV, the lepton and the $b$ quark will be within $\Delta R = 0.4$ of each other approximately 44/66\% of the time.%
\footnote{We obtained these numbers from chiral \ttbar\ resonance samples at parton-level using the TopBSM package in \MadGraph/\MadEvent\ {\tt 4.4.13} \cite{Alwall:2007st,Frederix:2007gi}.}  For $p_T \simeq 2$ TeV, this increases to 86/93\%.  Although the application of standard isolation criteria is a simple way to ensure reliability of the reconstructed leptons, the resulting low efficiency and polarization bias are major drawbacks.  Here we will explore what is possible using leptons that are non-isolated according to traditional measures.

Leptons found inside of jets have traditionally been considered unusable as independent objects.  Non-isolated leptons are produced in the decays of hadrons containing heavy quarks, and are in fact used as a standard heavy flavor tag.  Heavy flavor may be produced either promptly in the hard collision or from gluon splittings in the parton showers, the latter becoming progressively more common in light QCD jets at higher energies.%
\footnote{This suggests that some care must be taken if $b$-tags are ultimately used to discriminate semileptonic boosted top candidates.  QCD jets containing leptons are already enriched with heavy flavor.  Therefore the main utility of a $b$-tag on the semileptonic top would be to suppress the $W$-strahlung background.  As noted above, we will not explore $b$-tags here.}  In addition, there may also be contributions from decays of light mesons.  Instrumentation and material effects present further complications.

Electrons are particularly difficult to identify because they can look similar to $\pi^0$'s after accounting for electromagnetic showering in the inner tracking material.  It is not clear how difficult this discrimination will be in the crowded environment of a TeV-scale jet.  Understanding this issue requires detailed detector simulations and/or actual data, and therefore we defer investigation of electrons to the experimentalists (see, e.g.,~\cite{Brooijmans:2009boo}).  Hard muons inside of jets, on the other hand, are much less susceptible to instrumental fakes.  Moreover, we anticipate that any fake muons will be largely eliminated by our procedures outlined below for dealing with physics backgrounds.  Here we only consider backgrounds with real muons.

Muons coming from bottom and charm decays have a number of characteristics that distinguish them from muons originating from top decays.  The obvious difference is that the mass scale between the muon and its accompanying jet is controlled by $m_t$ in the case of top decay, whereas it may be much smaller for jets with heavy mesons.  In addition, there are discriminators which can be phrased more geometrically.  For a top decay at rest, the muon is largely uncorrelated in direction with the \bjet\ particles.  Consequently, while the muon and \bjet\ may end up close together in \DR\ in the lab frame, generally there is a gap between them.   This gap is characterized by the inverse of the boost, $O(m_t/p_{Tt})$.  The analogous quantity in heavy meson decay is $m_b/p_{Tb}$ or $m_c/p_{Tc}$, which is smaller unless the meson is relatively soft.  Another important difference can be found in the shower generated by the original hard parton.  For top, most of this radiation lives outside of the ``dead cone,'' again characterized by $\Delta R \sim m_t/p_{Tt}$.  The muon still typically remains isolated at this angular scale, even accounting for the additional radiation generated before the top's decay.  For bottom and charm quarks, the shower instead continues to the potentially much smaller angles $m_b/p_{Tb}$ and $m_c/p_{Tc}$.  Therefore, a muon produced in heavy meson decay will be sitting within a cloud of its sister decay products and particles produced in the preceding parton shower, whereas a muon produced in top decay will be approximately isolated out to \DR $\sim m_t/p_{Tt}$.

Several other useful observations about QCD background muons were made in \cite{Thaler:2008ju}:
\bit
\item  The fraction of the total visible jet energy carried by the muon, $z_\mu$, is typically small.  Most heavy quarks within TeV-scale jets are produced late in the shower, and then radiate further before hadronization.  On average, the muon then carries away approximately 1/3 of the already modest heavy quark energy.
\item  The muon is typically very well-aligned with the center of the ($b$-)jet: $\Delta R_{b\mu} \ll 1$.  This is simply due to the collinear enhancements within the parton shower, and the fact that the heavy mesons within the jet are highly boosted.  Indeed, criteria such as $\Delta R_{b\mu} > 0.4$ are common prerequisites for muon reconstruction.  However, for very high-$p_T$ top-jets, it may be useful to consider the effectiveness of cuts at smaller $\Delta R_{b\mu}$.
\item  Removing the muon from the jet typically results in a very small mass-drop,
\beq
x_\mu \equiv 1 - \frac{m_b^2}{m_{b\mu}^2},
\label{eq:xmu}
\eeq
where $m_b$ and $m_{b\mu}$ are the mass of the $b$-jet candidate with/without the muon included.  This is to some extent a combination of the previous two discriminators.  However, because $x_{\mu}$ uses the jet's mass, it is also sensitive to the distribution of the jet's consituents. For instance, even a relatively hard, wide-angle muon is usually accompanied by correlated hadrons, tending to push $x_{\mu}$ toward smaller values.
\eit
The authors of \cite{Thaler:2008ju} explored the discriminating power of these variables ($z_\mu$, $\Delta R_{b\mu}$, $x_\mu$) for jets initiated by prompt heavy flavor production, with visible mass of the muon-jet system, $m_{b\mu}$, above 100 GeV.  (They also studied the effect of varying an upper cut on $m_{b\mu}$, which we do not consider here.)

A priori, the best combination of cuts is not obvious.  Here, we study the performance of a handful of variables for discriminating generic QCD jets from genuine boosted semileptonic tops.  Our analysis includes the three variables of \cite{Thaler:2008ju} described above.  We also scan over the invariant mass $m_{b\mu}$, for which fixed cuts have been considered originally in \cite{Agashe:2006hk}, and also in \cite{Thaler:2008ju}.  Finally, we consider a novel ``mini-isolation'' cut at tracker level inspired by the geometric observations above, and which we now describe.

To form an isolation variable, we must define a cone size.  In top decay, the separation between the muon and \bjet\ scales inversely with the top-jet $p_T$, suggesting that we take a cone size with this scaling.  Alternately, we may attempt to capture the decay products of a hypothetical heavy flavor parent, which ideally would use a cone scaling inversely with the heavy meson momentum.  More realistically, we can use the $p_T$ of the muon itself as a rough tracer.  We found that the latter choice results in $O(1)$ better discrimination, since it also acts in part like a cut on the muon hardness.  Softer muons are given larger isolation cones, and are consequently more difficult to isolate.  Specifically, we find that a cone size 
\beq
R_{iso} = \frac{15\;{\rm GeV}}{p_{T\mu}} \simeq \frac{3m_B}{p_{T\mu}}
\label{eq:Riso}
\eeq
works well, and we take this to be our nominal cone definition.  From this, we define an isolation variable
\beq
mini\mbox{-}iso \equiv \frac{p_{T\mu}}{p_{T{\rm cone}}},
\label{eq:mini-iso}
\eeq
where the denominator scalar-sums the $p_T$s of all charged particles with $p_T > 1$ GeV in the cone, including the muon.\footnote{The 1 GeV tracking cutoff roughly models the critical $p_T$ for spiral-out in the ATLAS and CMS magnetic fields.}  We emphasize that this specific choice has been only very coarsely optimized, and that finding the best cone merits further study under more realistic conditions.%
\footnote{It will also be important to determine at what point precision tracking actually breaks down due to crowding of hits.  It is possible that some of the muons removed by our isolation cut under idealized conditions can be rejected in the real experiments simply due to poor tracking in the inner detectors.  For muons from genuine boosted top decays, which tend to be better isolated, we expect that this tracking breakdown will be much less likely.}

\subsection{Event simulation and reconstruction}

\label{subsec:reco}

To get an estimate of the production of muons through QCD processes, we study generic dijet event samples generated with \PYTHIA\ {\tt 6.4.15}~\cite{Sjostrand:2006za} and \HERWIG\ {\tt 6.510}~\cite{Corcella:2000bw}, with default settings in 14 GeV $pp$ collisions.  The simulations implicitly include both prompt and radiative production of heavy flavor, and the \PYTHIA\ samples also include decays-in-flight of light mesons.%
\footnote{We allow particles to decay within a cylindrical volume of half-length 4m and radius 2m.}  We do not assume that these represent fully trustworth representations of the physics in this untested energy region, but we will see shortly that the effectiveness of our cuts allows us a large margin of error.

We compare these to signal samples consisting of decayed color-octet spin-1 bosons with pure left- and right-chirality couplings to tops, generated with the TopBSM package of \MadGraph/\MadEvent\ {\tt 4.4.13} (+\PYTHIA)~\cite{Alwall:2007st,Frederix:2007gi}.   For completeness, we also include $Wjj$ (``$W$-strahlung'') simulations, similarly generated with \MadGraph/\MadEvent, which will be described in more detail in the next section.

Reconstruction of the simulated events is similar to \cite{Kaplan:2008ie}, with several modifications.  We first demand the presence of at least one muon with $p_T > 30$ GeV and $|\eta| < 2.5$.  We set the leading muon aside, and then deposit all other particles into an idealized calorimeter consisting of perfect energy-sampling cells of size $\Delta\eta\times\Delta\phi = 0.1\times0.1$.  (We do not apply a magnetic field.)  We then cluster the cells using the Cambridge/Aachen algorithm implemented in \FastJet\ {\tt 2.3.4}~\cite{Cacciari:2005hq}.  The clustering radius is picked according to the event's $H_T$ (scalar-summed $p_T$) measured in the semicylindrical region opposite the muon in $\phi$, and with $|\eta| < 3.0$:  $R = \{0.8,0.6,0.4\}$ for $H_T$ $> \{500,800,1300\}$ GeV.%
\footnote{It is also possible to use a fixed ``fat'' clustering scale, but somewhat greater care will be required in order to remove jet activity uncorrelated with the top decays, particulary FSR off of the top itself.  Our variable clustering radius exploits the fact that the angular separation between top decay products shrinks as the event energy scale increases.  (For other methodology and uses of variable jet clustering radius, see \cite{Krohn:2009zg}.)  We also note that the \bjet\ in semileptonic decays may be reclustered using a somewhat smaller scale in order to improve momentum and mass resolution, but we have not explored this.}  We only keep jets which are above $p_T = 50$ GeV and with $|\eta| < 2.5$.  

Next, we identify the candidate hadronic top-jet and \bjet.  The former is simply taken to be the highest-$p_T$ jet in the event.  We require this jet to be above $p_T = 500$ GeV, and that it live in the central part of the calorimeter, $|\eta| < 1.5$, where the granularity is finest in both CMS~\cite{Bayatian:2006zz} and ATLAS~\cite{:1999fq}.  This latter requirement focuses our attention on those top-jet candidates that are most suitable for top-tagging using the calorimeter, but in any case these jets should have the best mass resolution for simpler mass-based tags.  The \bjet\ we identify as the remaining jet closest to the leading muon, without application of an explicit $b$ tag.  The semileptonic top candidate is then the sum of this \bjet\ candidate and the leading muon, with \met\ folded in according to some prescription.  However, we will attempt to get as far as possible without using \met\ explicitly, postponing its introduction until section \ref{sec:resonances}.  One consequence of this is that whenever we refer to the $p_T$ of the semileptonic top candidate below, we will actually be using the $p_T$ of the recoiling hadronic top candidate as a proxy.

At this level of analysis, we reject about 99\% of light quark jets and 97$\sim$98\% of gluon jets, simply from the requirement of the hard muon.  The exact numbers depend somewhat on $p_T$, as well as on the simulation.  In particular, \PYTHIA\ appears to have higher pass rates, by a factor of about $1.5$.  Prompt heavy-flavor jets of course pass with much higher efficiency, essentially determined by their branching fractions to muons.

\subsection{Discriminator analysis and choice of nominal cuts}

\label{subsec:hfcuts}

\begin{figure}[tp]
\begin{center}
\epsfxsize=0.44\textwidth\epsfbox{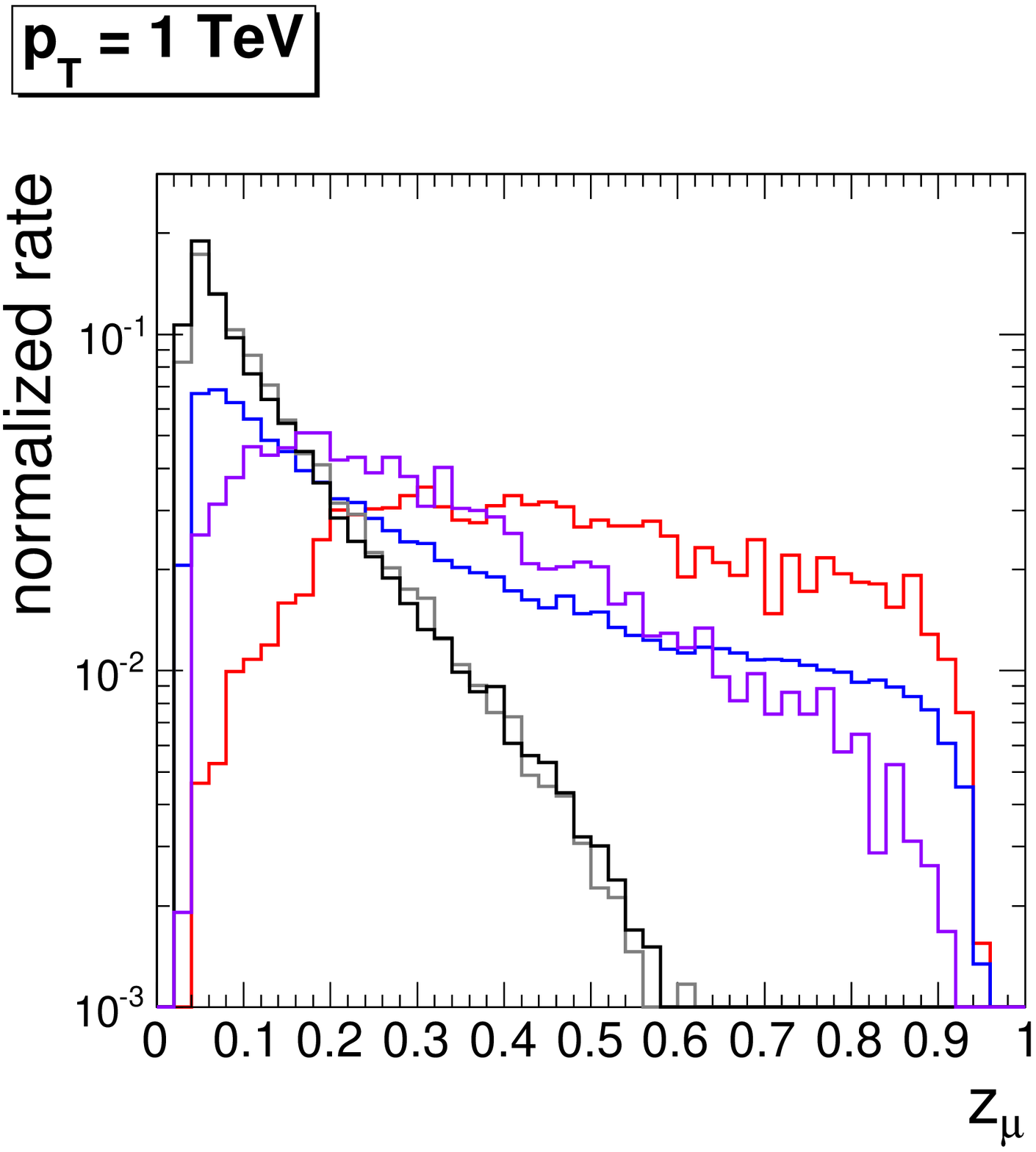}
\epsfxsize=0.44\textwidth\epsfbox{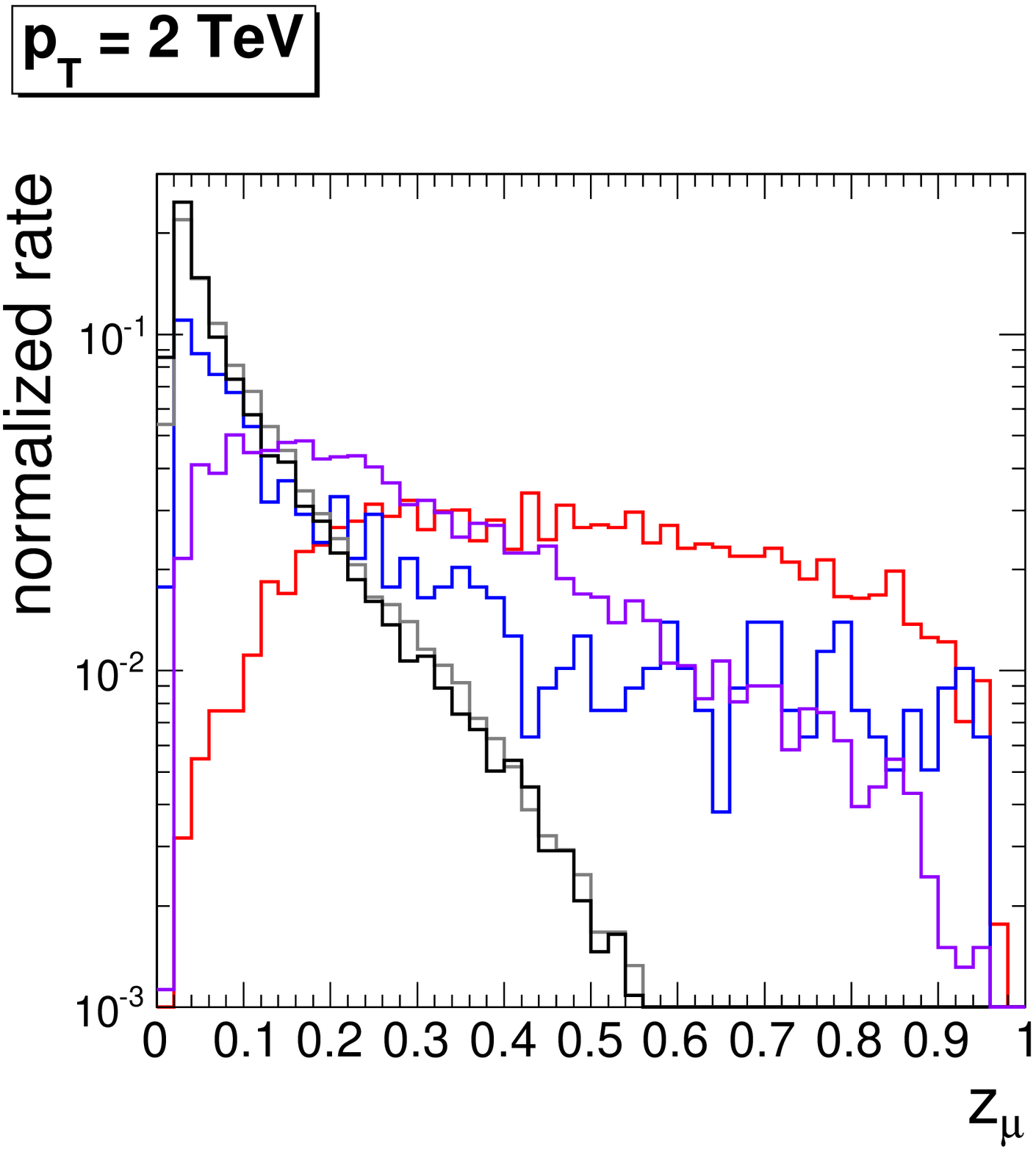}
\caption{Distributions of the muon energy fraction $z_\mu$ for 1 TeV and 2 TeV semileptonic top candidates reconstructed from LH chiral \ttbar\ (purple), RH chiral \ttbar\ (red), PYTHIA QCD (black), HERWIG QCD (grey), and $Wjj$ (blue).}
\label{fig:zmu}
\end{center}
\end{figure}

\begin{figure}[tp]
\begin{center}
\epsfxsize=0.44\textwidth\epsfbox{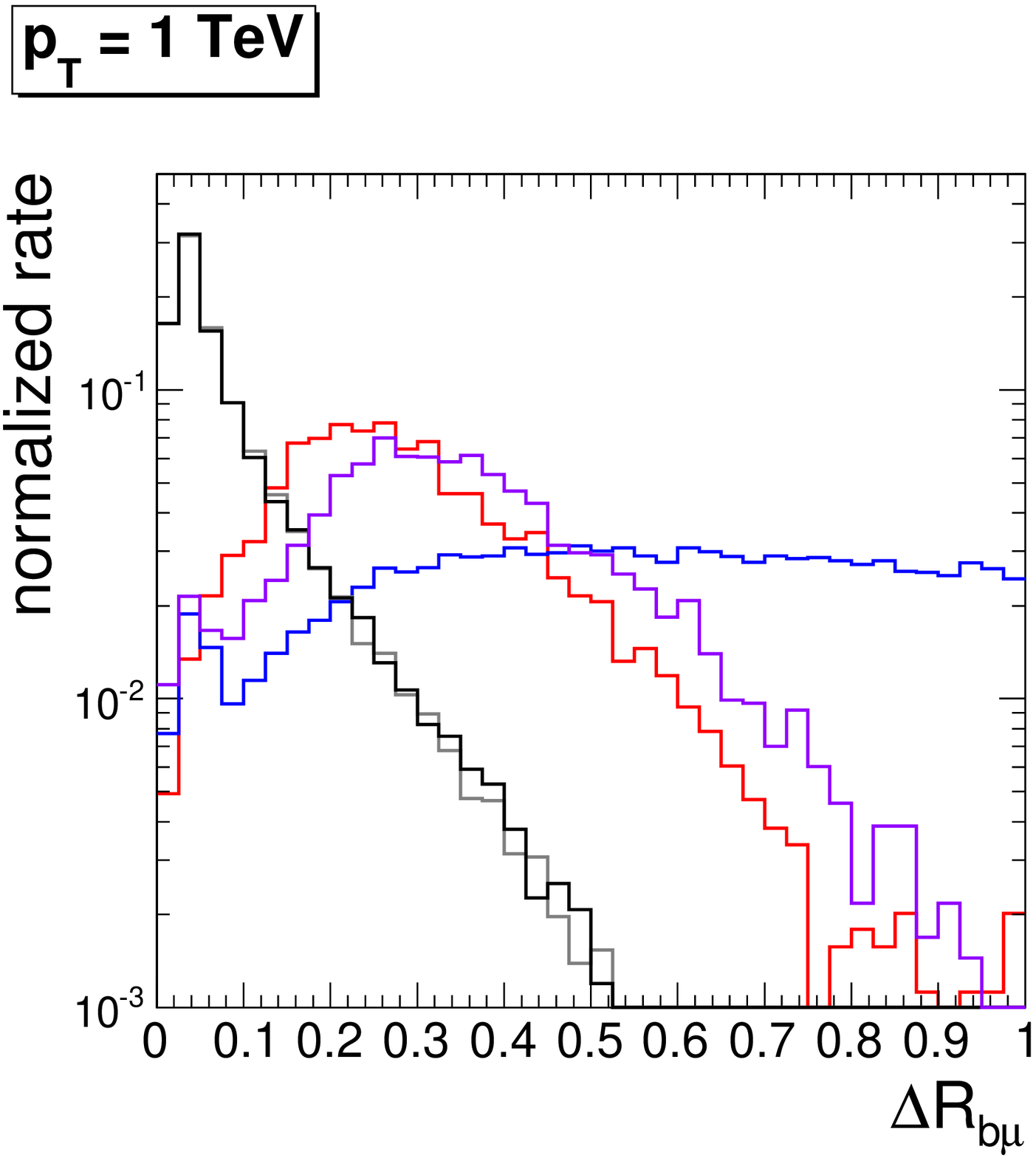}
\epsfxsize=0.44\textwidth\epsfbox{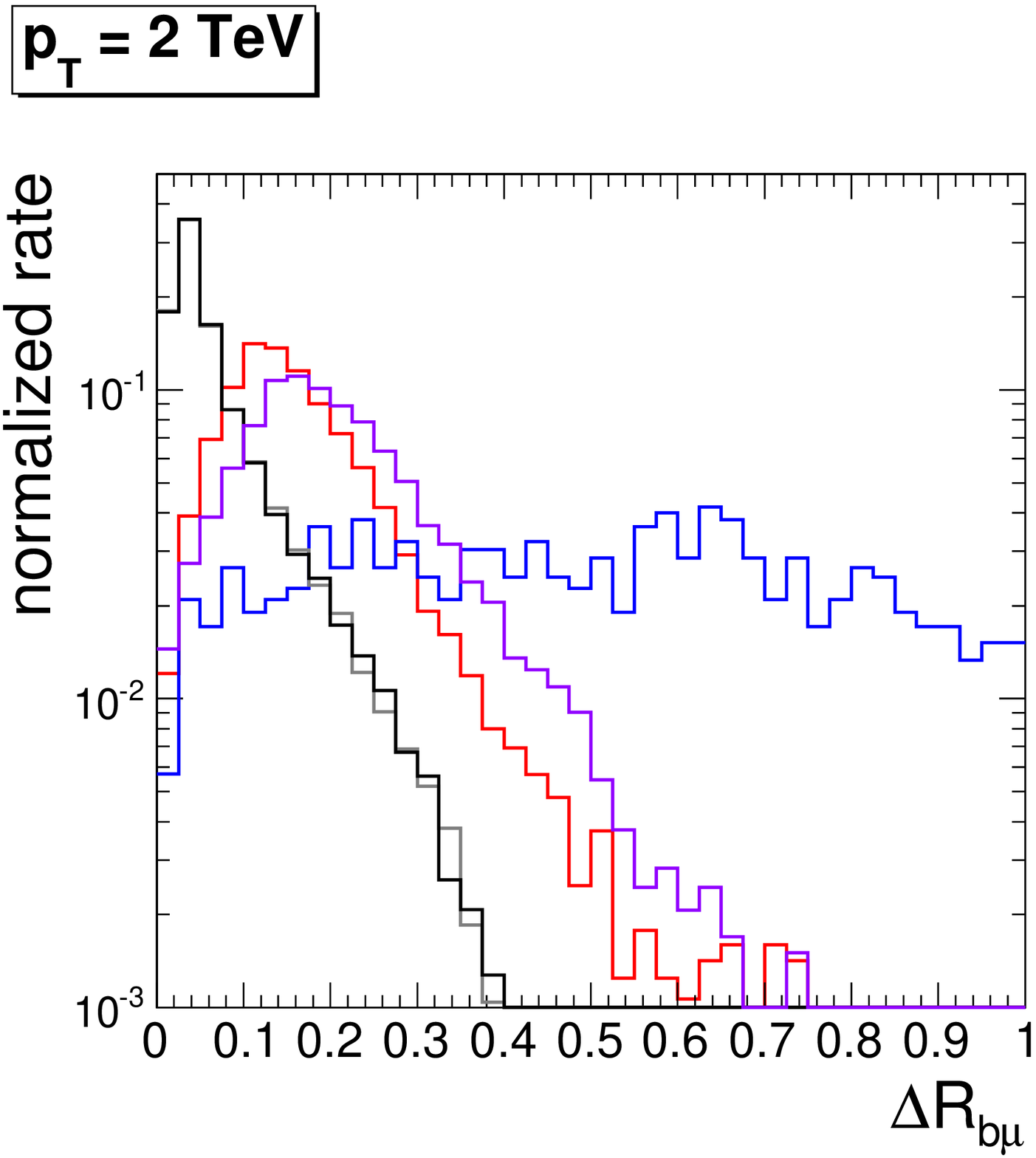}
\caption{Distributions of $\Delta R_{b\mu}$  for 1 TeV and 2 TeV semileptonic top candidates reconstructed from LH chiral \ttbar\ (purple), RH chiral \ttbar\ (red), PYTHIA QCD (black), HERWIG QCD (grey), and $Wjj$ (blue).}
\label{fig:DRbl}
\end{center}
\end{figure}

\begin{figure}[tp]
\begin{center}
\epsfxsize=0.44\textwidth\epsfbox{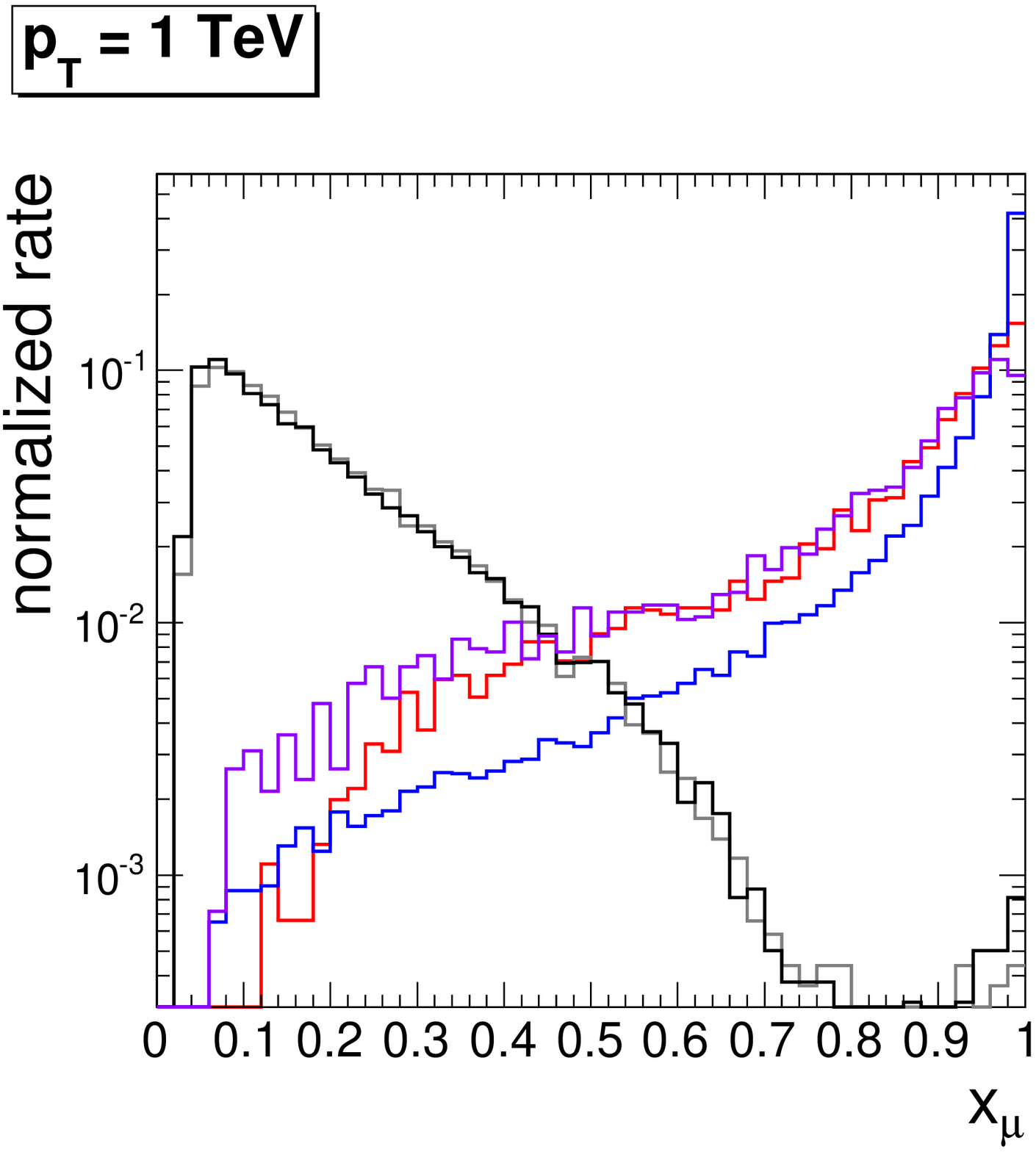}
\epsfxsize=0.44\textwidth\epsfbox{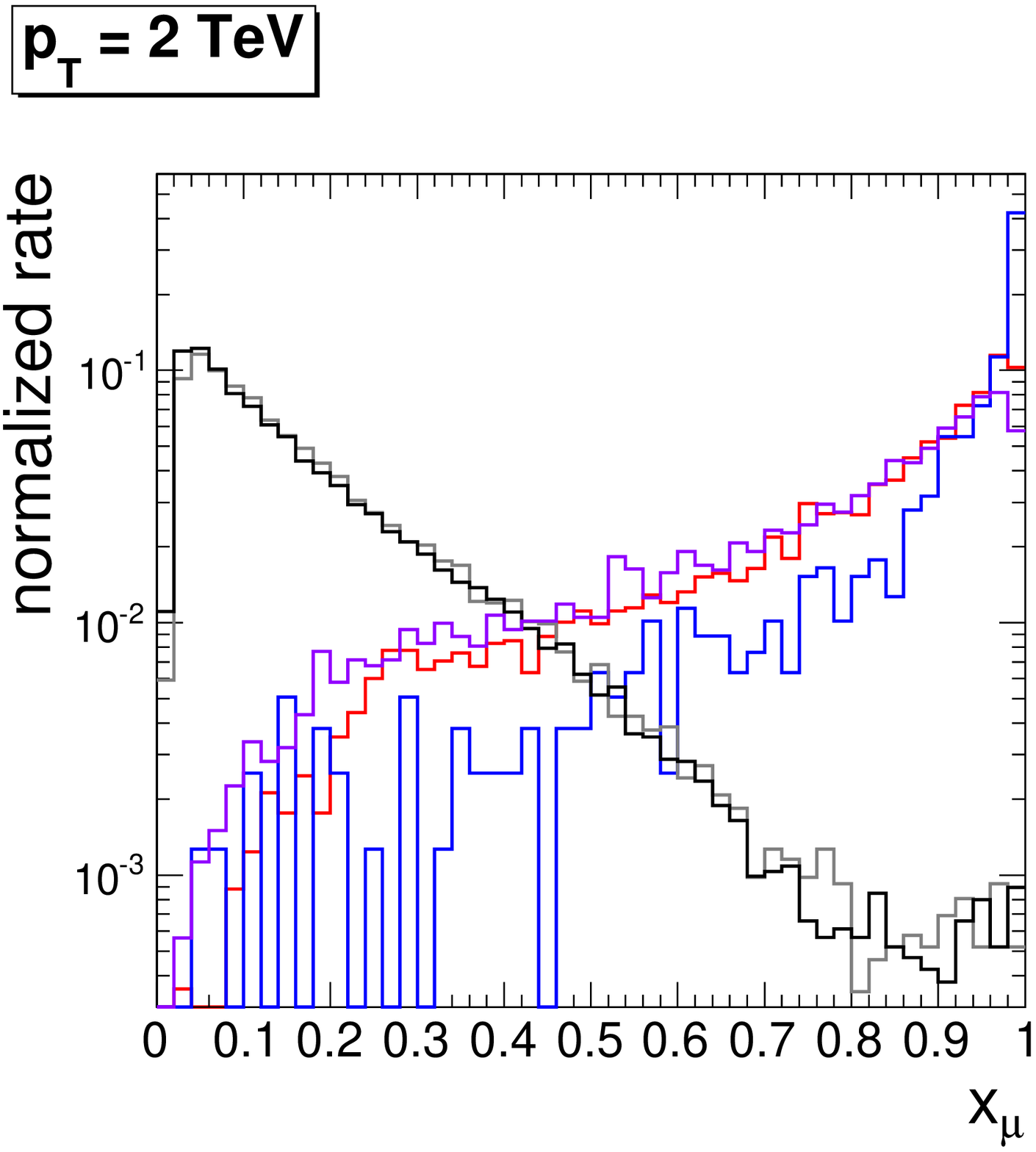}
\caption{Distributions of the mass-drop $x_\mu$ (Eq.~\ref{eq:xmu}) for 1 TeV and 2 TeV semileptonic top candidates reconstructed from LH chiral \ttbar\ (purple), RH chiral \ttbar\ (red), PYTHIA QCD (black), HERWIG QCD (grey), and $Wjj$ (blue).}
\label{fig:xmu}
\end{center}
\end{figure}

\begin{figure}[tp]
\begin{center}
\epsfxsize=0.44\textwidth\epsfbox{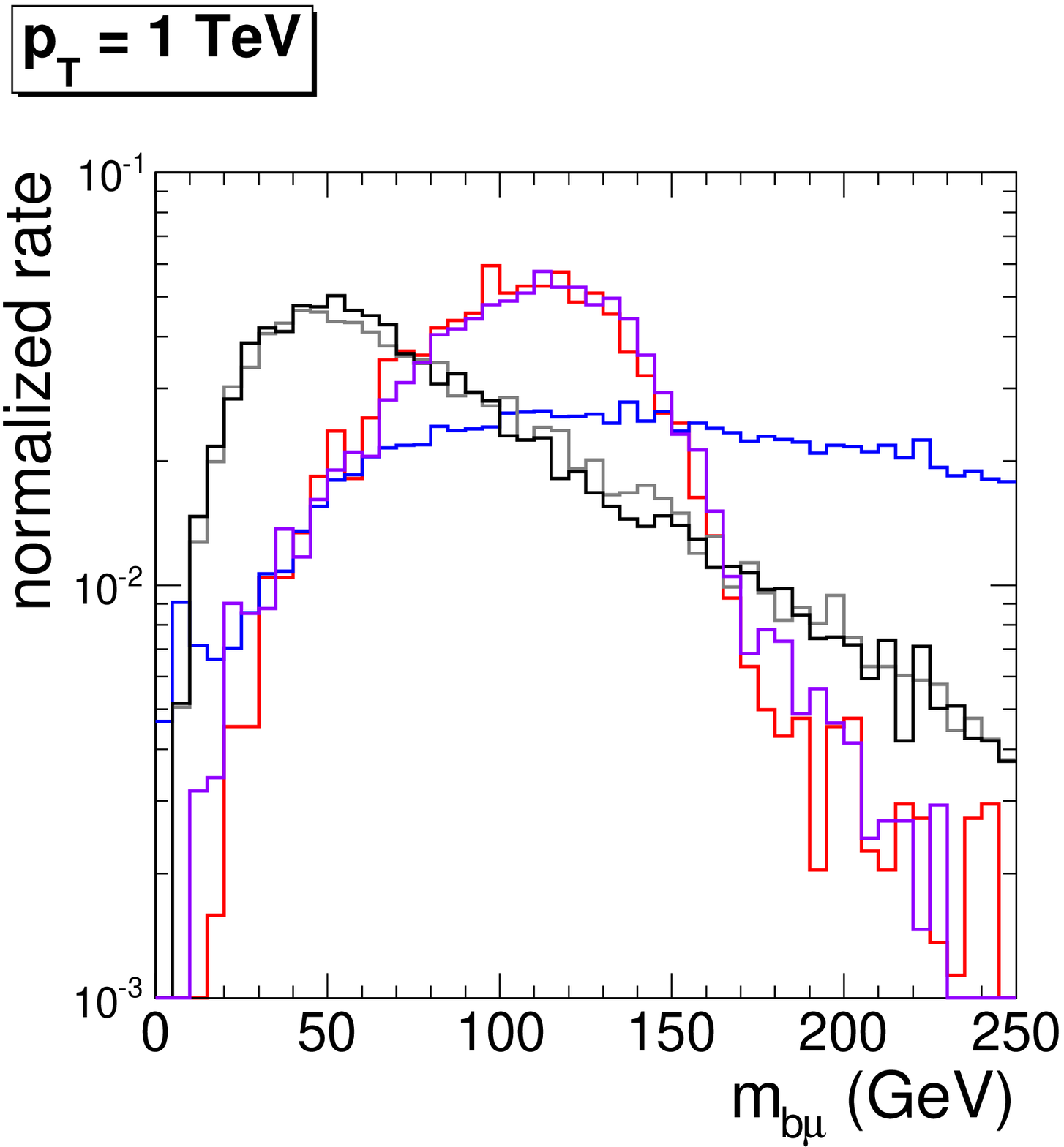}
\epsfxsize=0.44\textwidth\epsfbox{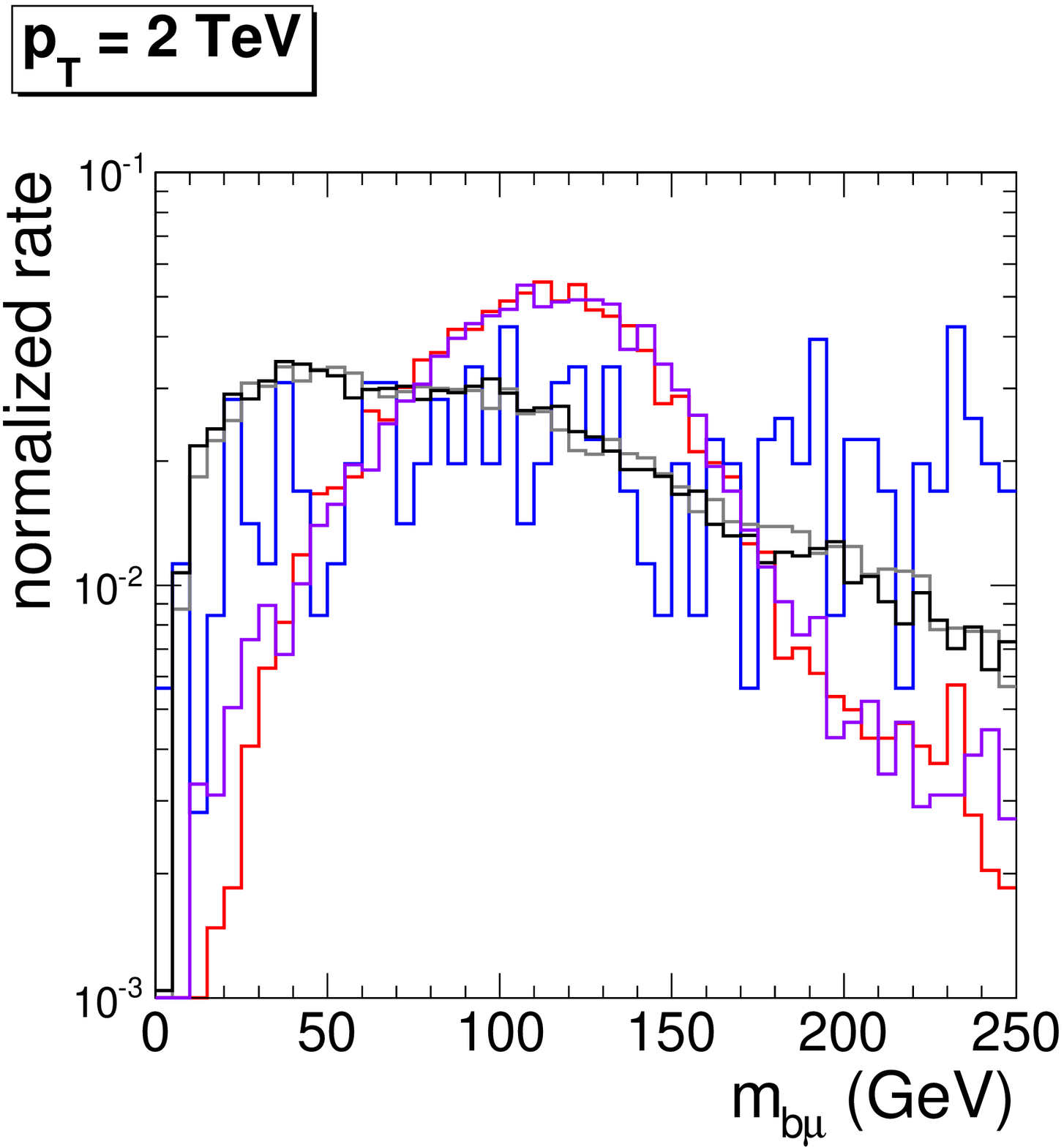}
\caption{Distributions of $m_{b\mu}$  for 1 TeV and 2 TeV semileptonic top candidates reconstructed from LH chiral \ttbar\ (purple), RH chiral \ttbar\ (red), PYTHIA QCD (black), HERWIG QCD (grey), and $Wjj$ (blue).}
\label{fig:mbl}
\end{center}
\end{figure}

\begin{figure}[tp]
\begin{center}
\epsfxsize=0.44\textwidth\epsfbox{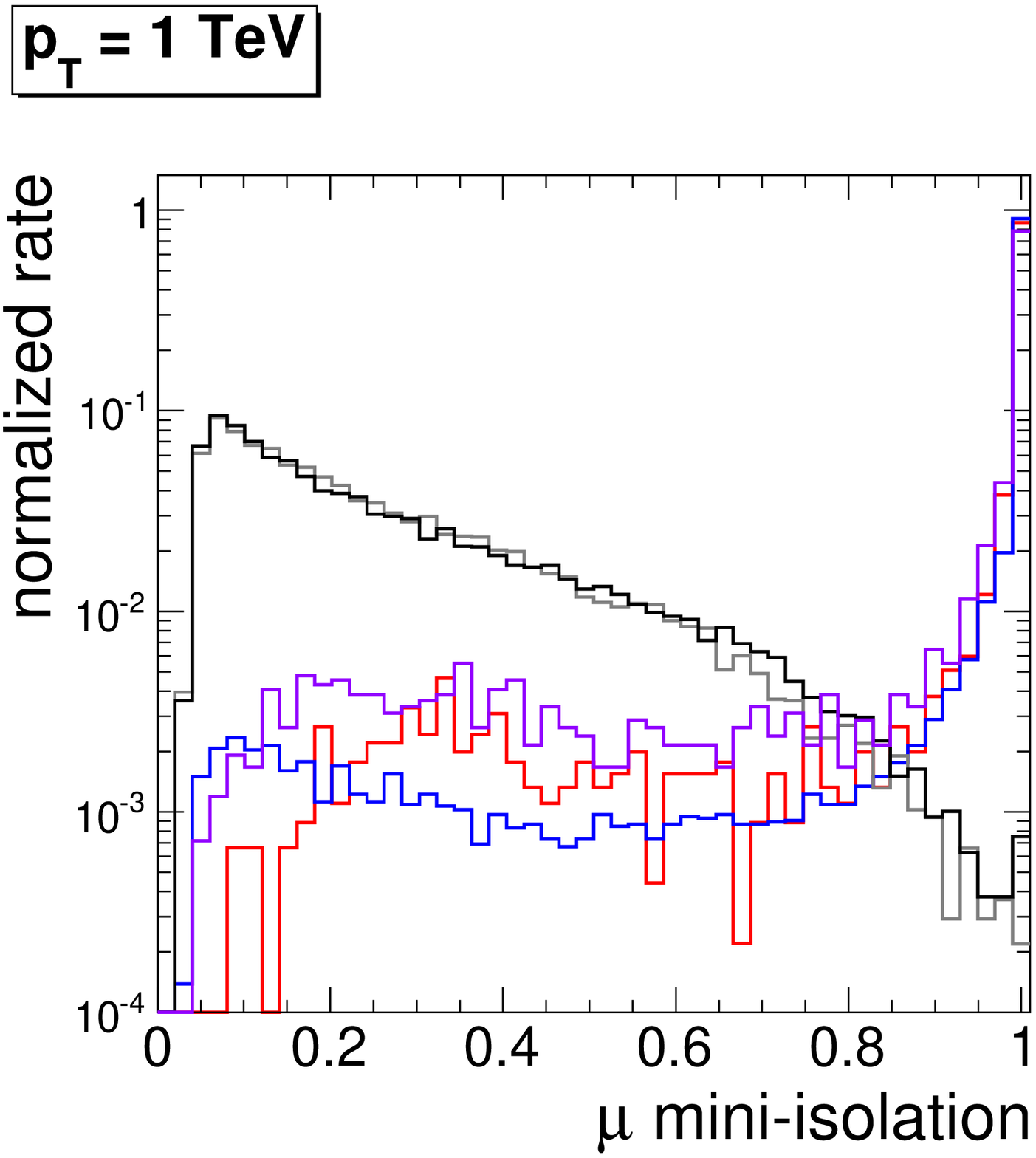}
\epsfxsize=0.44\textwidth\epsfbox{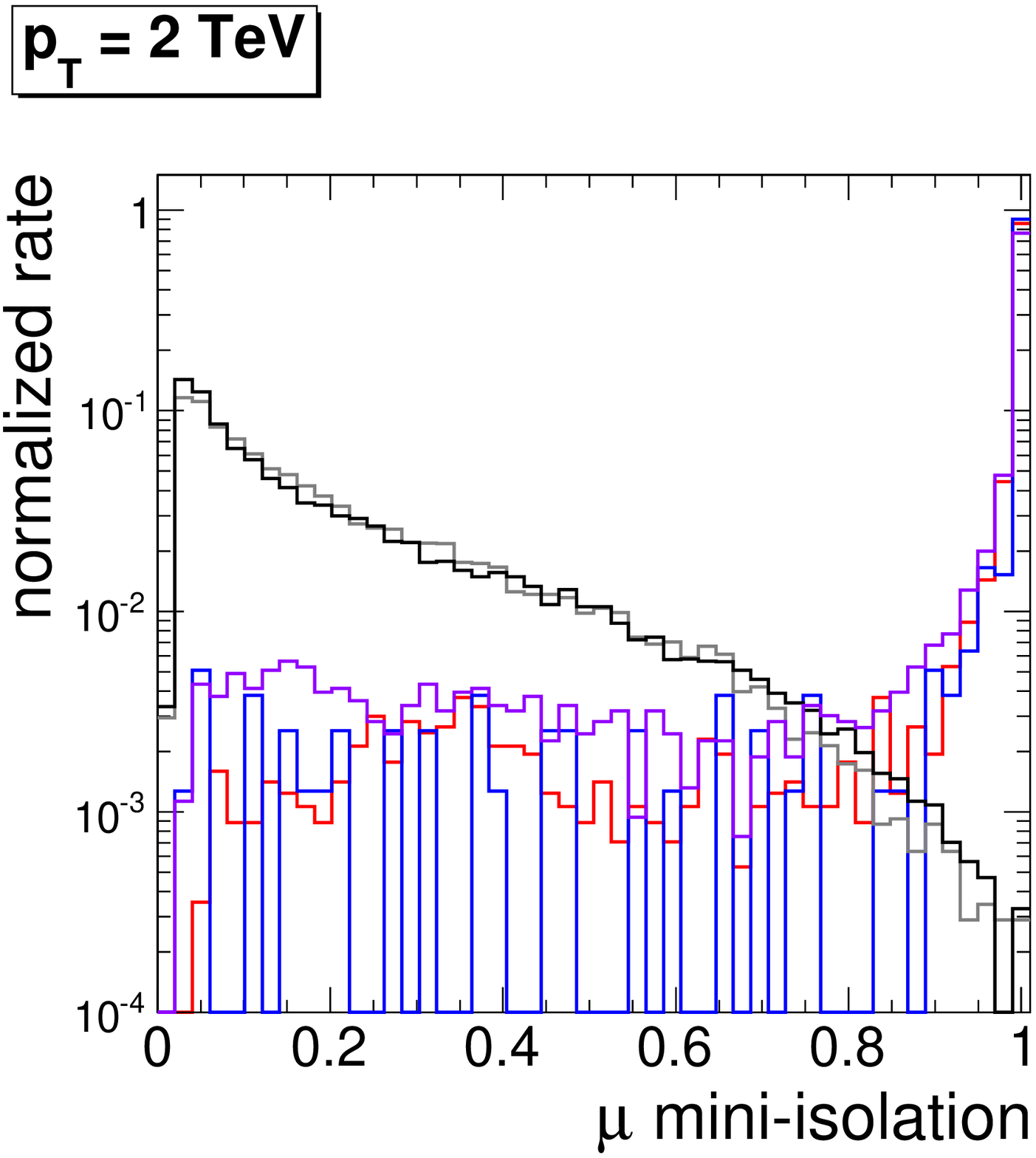}
\caption{Distributions of the muon mini-isolation (Eqs.~\ref{eq:Riso} and~\ref{eq:mini-iso}) for 1 TeV and 2 TeV semileptonic top candidates reconstructed from LH chiral \ttbar\ (purple), RH chiral \ttbar\ (red), PYTHIA QCD (black), HERWIG QCD (grey), and $Wjj$ (blue).}
\label{fig:iso}
\end{center}
\end{figure}

\begin{figure}[tp]
\begin{center}
\epsfxsize=0.44\textwidth\epsfbox{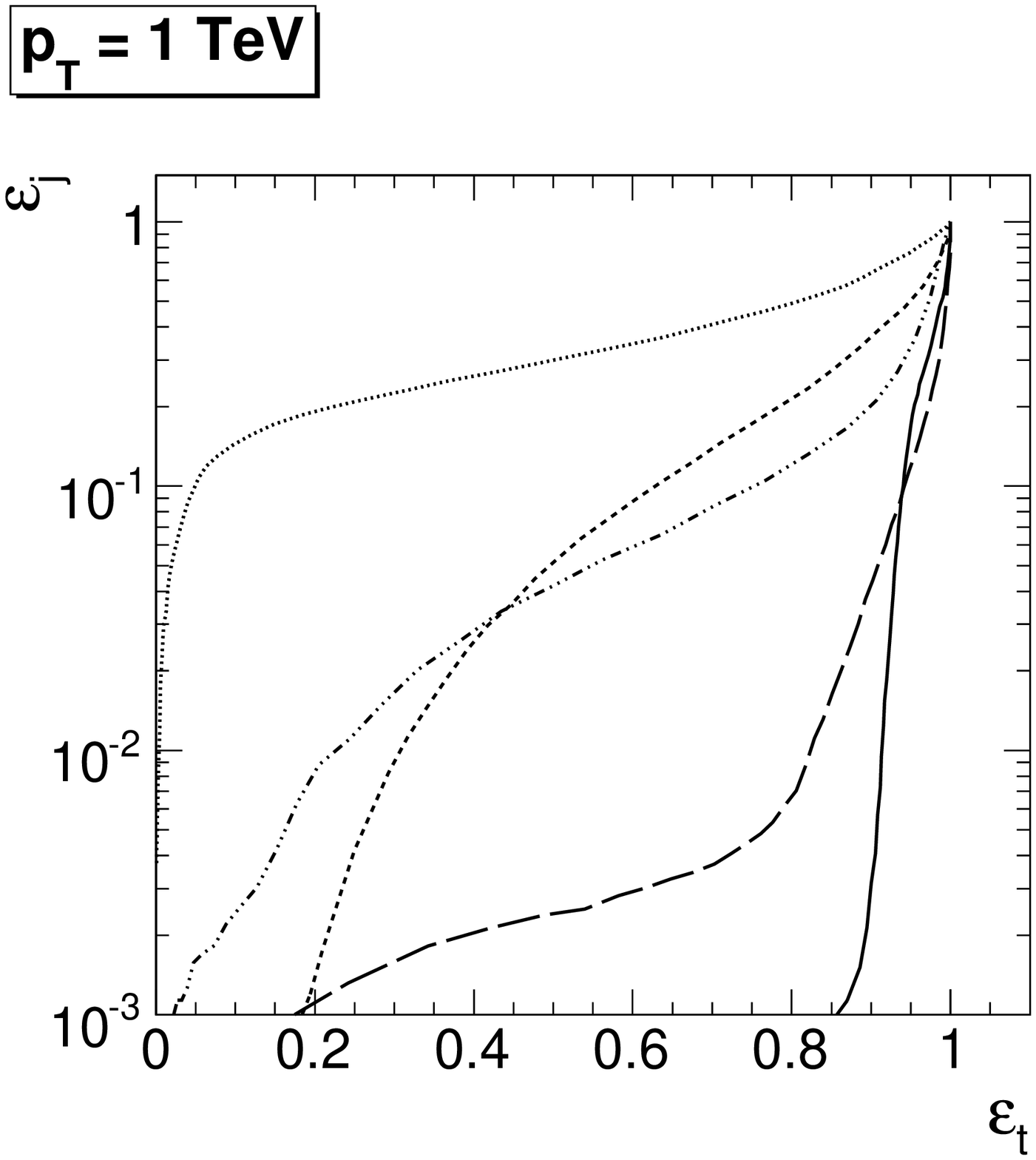}
\epsfxsize=0.44\textwidth\epsfbox{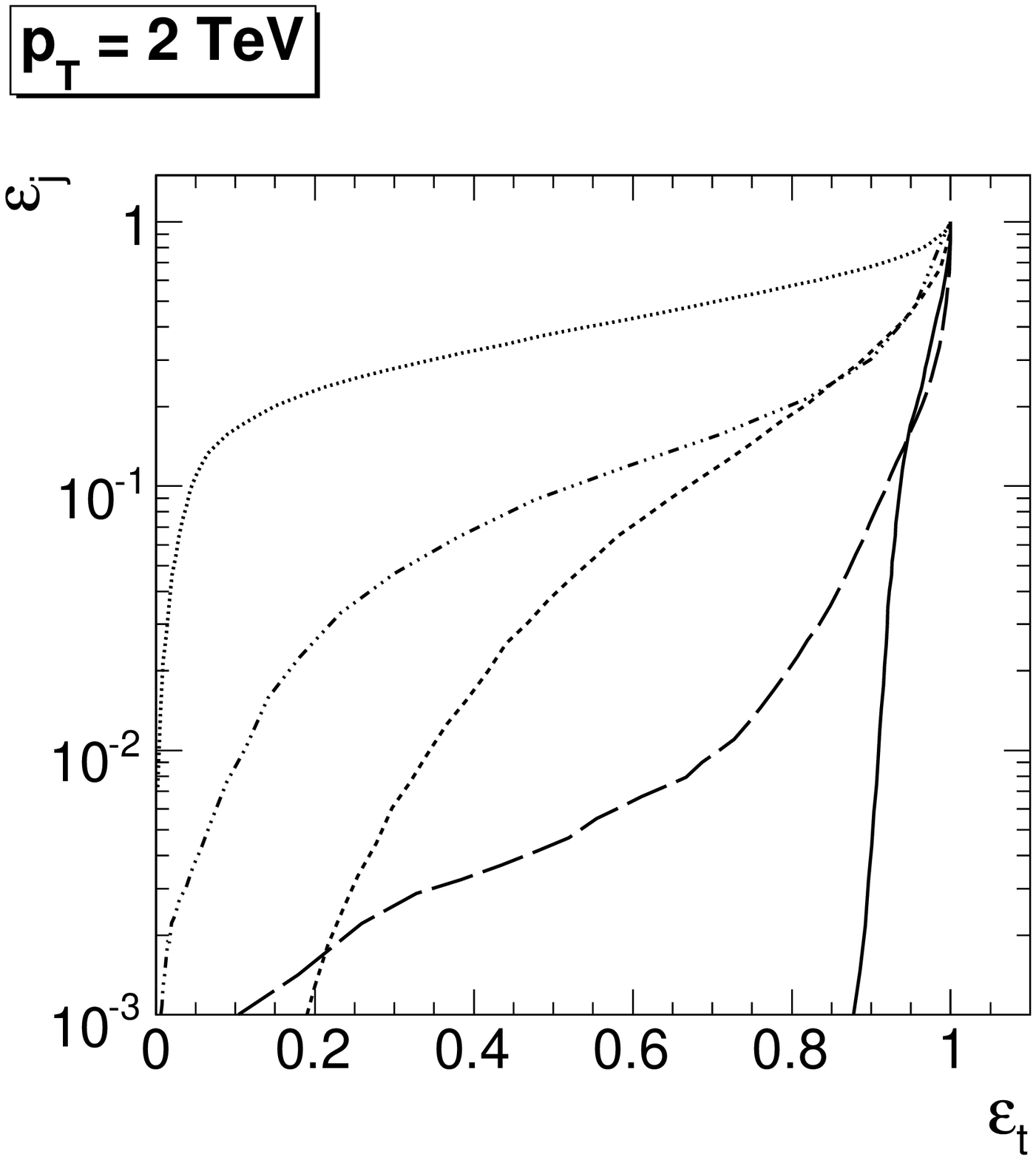}
\caption{Efficiency for PYTHIA QCD-jets (reconstructed as candidate semileptonic top-jets) versus efficiency for helicity-averaged top-jets at 1 TeV and 2 TeV, scanning over independent one-sided cuts described in the text.  The variables used are $z_\mu$ (short dash), $\Delta R_{b\mu}$ (dash-dots), $x_\mu$ (long dash), $m_{b\mu}$ (dots), and $mini$-$iso$ (solid).}
\label{fig:epsilonSB}
\end{center}
\end{figure}

To get an idea of how best to discriminate semileptonic boosted tops from light jets, we scan over signal and background efficiencies obtained by independent 1D, one-sided cuts on the five variables discussed in subsection~\ref{subsec:lepsinjets}: $z_\mu$, $\Delta R_{b\mu}$, $x_\mu$, $m_{b\mu}$, and $mini$-$iso$.  To get a sense for how the effects of these cuts scale with top $p_T$, we look at candidate semileptonic top-jets at $p_T \simeq 1$ TeV and at $p_T \simeq 2$ TeV.  The normalized distributions for these variables are shown in Figs.~\ref{fig:zmu} to \ref{fig:iso}, and their helicity-averaged discrimination curves using \PYTHIA\ dijets in Fig.~\ref{fig:epsilonSB}.  The discrimination curves for \HERWIG\ are very similar, as can be inferred from the distributions of the individual variables.

In this analysis, for sake of brevity, we have not individually distinguished light quark, gluon, and prompt heavy-flavor jets within the continuum dijet simulations described above.  The results will therefore most directly apply to searches in \ttbar, where dijets are the relevant QCD background.  We note that the composition of jets analyzed is approximately 14\% light quarks, 56\% gluons, 18\% prompt $b$, and 12\% prompt $c$ for the 1 TeV \PYTHIA\ samples.  The composition shifts to 29\% light quarks, 51\% gluons, 11\% prompt $b$, and 9\% prompt $c$ for the 2 TeV \PYTHIA\ samples.  The \HERWIG\ samples have similar composition.

From Fig.~\ref{fig:epsilonSB}, we infer that $x_\mu$ and $mini$-$iso$ are the best individual discriminating variables, with the latter displaying the strongest discrimination.  In fact, $mini$-$iso$ appears to be such a good discriminator that we can apparently drive the background efficiency below the part-per-mil level while retaining more than 80\% of the signal.%
\footnote{We note that this conclusion is independent of the hard parton which initiates the QCD jet.}  However, it is far from clear that our modelling of the high $mini$-$iso$ tail of the QCD jets is so accurate that we can fully trust this prediction.  While the remarkably good agreement between \PYTHIA\ and \HERWIG\ in Fig.~\ref{fig:iso} is at least an encouraging indication that the physics might be under control, detector effects could also be very important.  In particular, some of the tracks in the immediate vicinity of the muon could be missed or misreconstructed, making a QCD-induced muon look more isolated than it is in reality.  We do not have the tools to address this issue here.  Nonetheless, as is clear from Fig.~\ref{fig:epsilonSB}, if even a fraction of $mini$-$iso$'s discriminating power survives in the full detector, it will remain the strongest variable.

In section \ref{sec:resonances}, we will perform a study of the \ttbar\ resonance search reach of the LHC, and for this we need to choose a nominal $mini$-$iso$ cut.  Given our uncertainty above, we do not push this cut as aggressively as we could, but instead we use the somewhat conservative choice of $mini$-$iso > 0.9$.  The efficiencies of this cut are given in the first row of Tables \ref{tab:1TeVdijet} and \ref{tab:2TeVdijet}.  We find a few part-per-mil acceptance of QCD jets, with only about 10\% loss of top-jet signal.  This by itself is enough to bring the QCD backgrounds to the resonance search under good control.

\begin{figure}[tp]
\begin{center}
\epsfxsize=0.44\textwidth\epsfbox{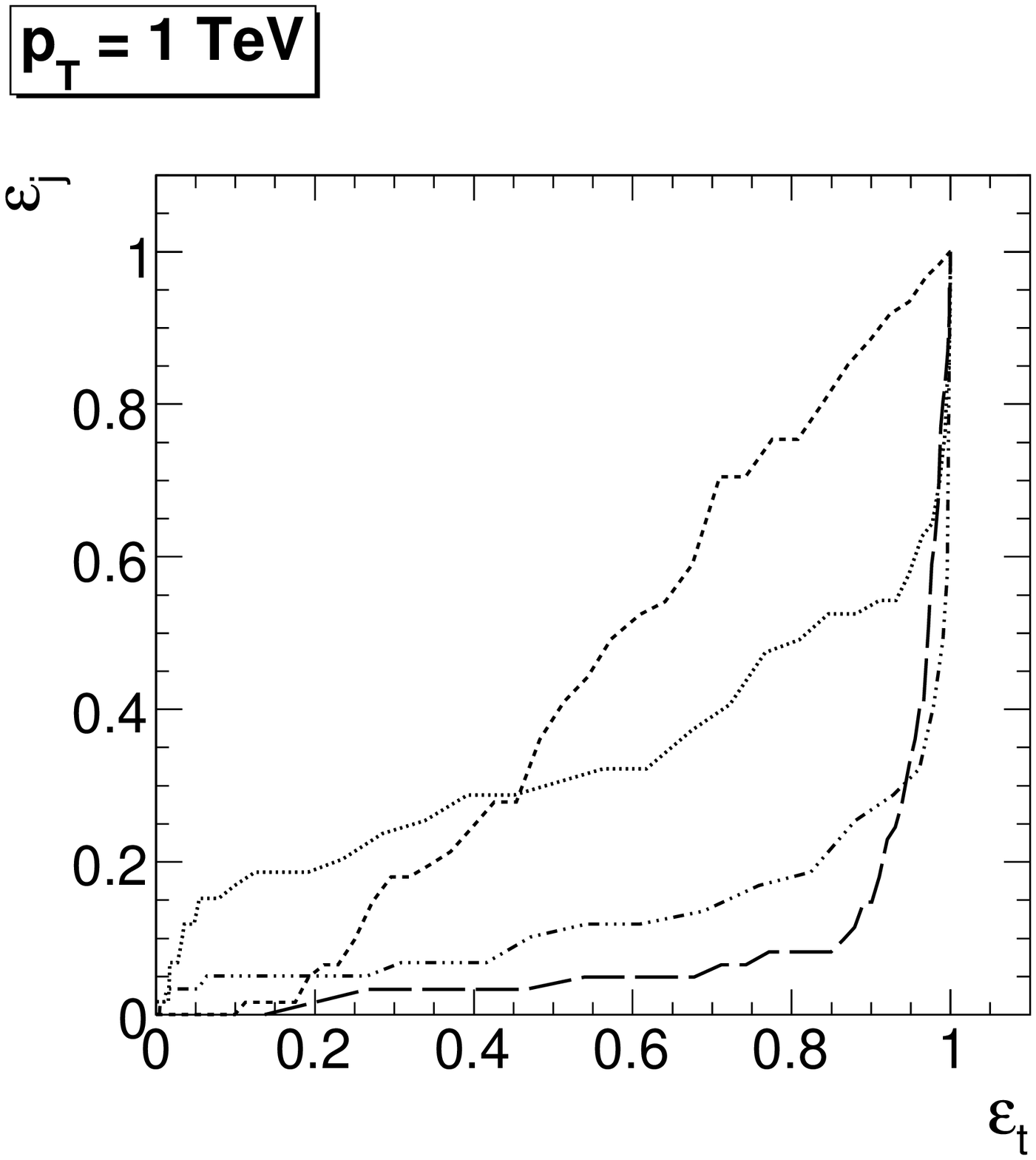}
\epsfxsize=0.44\textwidth\epsfbox{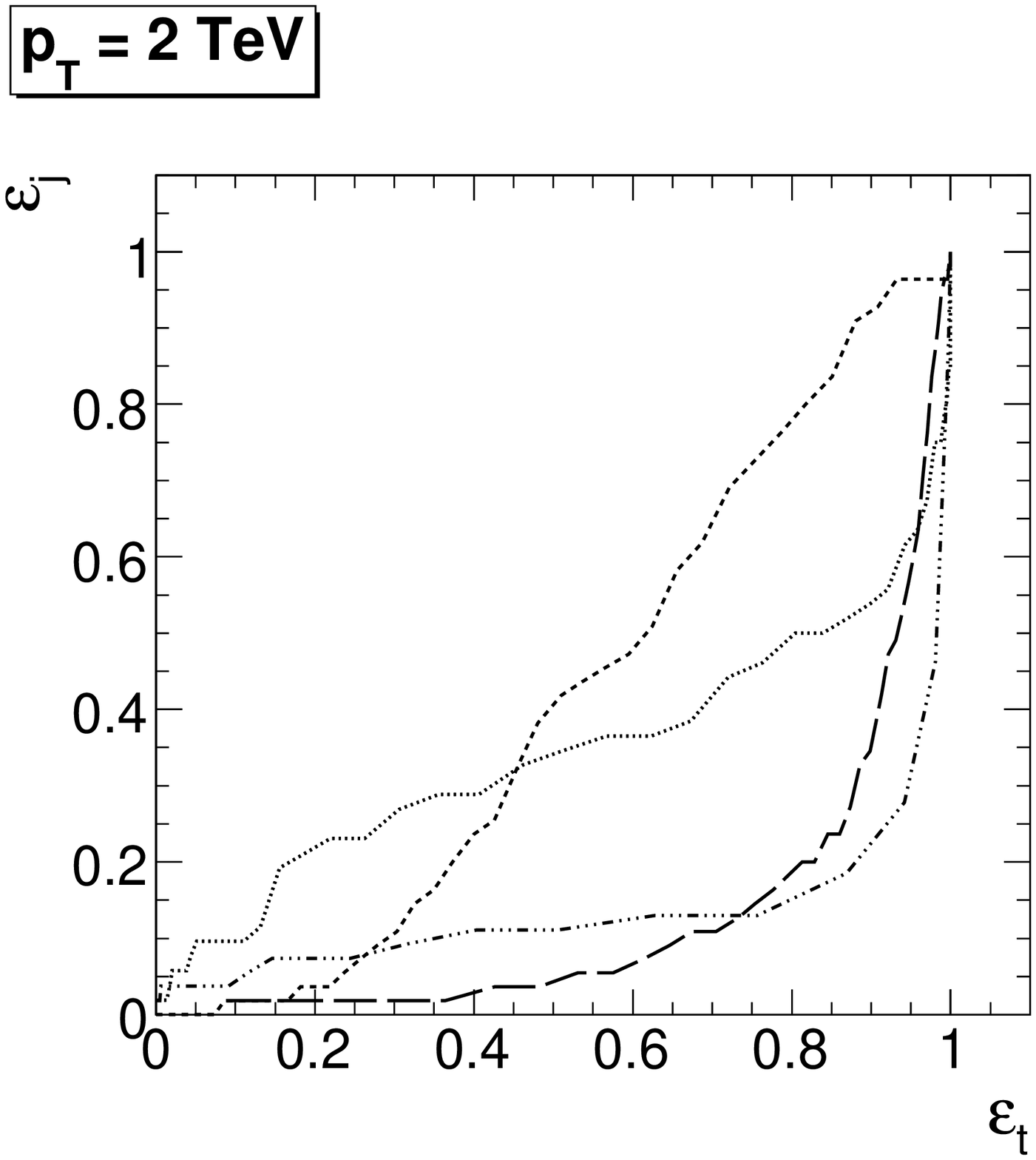}
\caption{Efficiency for PYTHIA QCD-jets (reconstructed as candidate semileptonic top-jets) versus efficiency for helicity-averaged top-jets, after the $mini$-$iso$ cut, at 1 TeV and 2 TeV, scanning over independent one-sided cuts described in the text.  The variables used are $z_\mu$ (short dash), $\Delta R_{b\mu}$ (dash-dots), $x_\mu$ (long dash), $m_{b\mu}$ (dots).  The fluctuations in the plot are due to depleted statistics in the QCD samples.}
\label{fig:epsilonSB.postIso}
\end{center}
\end{figure}

Even if the performance of $mini$-$iso$ proves to be less than ideal in the full experimental environment, we still have several other discriminating variables with which it can be combined.  To get a sense for how such a combination would perform, we explore the discriminating power of the remaining variables after application of the $mini$-$iso$ cut at $0.9$.  The helicity-averaged discrimination curves are displayed in Fig.~\ref{fig:epsilonSB.postIso}.

We see that $\Delta R_{b\mu}$ and $x_\mu$ are the best remaining variables.  Though $\Delta R_{b\mu}$ outperforms $x_\mu$ for the highest signal acceptance, the latter ultimately appears to allow greater discrimination.  For either variable, we find that we can achieve an additional factor of $2\sim 3$ reduction in the QCD efficiency with only a few percent loss of signal.  To take advantage of this, we will use a nominal cut of $x_\mu > 0.5$, though we also note that a cut of $\Delta R_{b\mu} > 0.10$ at 1 TeV and $\Delta R_{b\mu} > 0.05$ at 2 TeV would perform comparably.  The results are summarized in Tables \ref{tab:1TeVdijet} and \ref{tab:2TeVdijet}, first without the $mini$-$iso$ cut, and then in combination.%
\footnote{We obtain comparable efficiencies for QCD jets originating from different species of hard partons.}  In order to test the sensitivity of these numbers to our particular choice of $mini$-$iso$ cut, we also tried a looser cut of $mini$-$iso$ $> 0.75$.  We find that the signal efficiency is largely unaffected, and the background efficiency rises by a factor of about 6.  This is still a factor of 4 better than $x_\mu$ by itself.

In summary, in this subsection we have performed a very simple discriminator analysis for QCD jets with embedded muons versus boosted semi-muonic tops, using the five variables $z_\mu$, $\Delta R_{b\mu}$, $x_\mu$, $m_{b\mu}$, and $mini$-$iso$.  We find that the $mini$-$iso$ variable offers the most promise, but can also be fruitfully combined with $x_\mu$ or $\Delta R_{b\mu}$ for a factor of $2\sim 3$ improvement.  We propose a baseline set of cuts of 
\beq
mini\mbox{-}iso > 0.9, \;\;\;\; x_\mu > 0.5,  \label{eq:nominalhf}
\eeq
which can potentially achieve part-per-mil background efficiency with less than 20\% loss of signal, as indicated in Tables \ref{tab:1TeVdijet} and \ref{tab:2TeVdijet}.  In section \ref{sec:resonances}, we will see that these cuts can effectively eliminate the QCD backgrounds to the \ttbar\ invariant mass spectrum.

\begin{table}[t]
\caption{Nominal efficiencies at 1 TeV.}
\begin{center}
\begin{tabular}{ | c | c | c | c | c | c | c }
  \hline
                                                  & $t_L$ & $t_R$ &  \PYTHIA\ dijet & \HERWIG\ dijet & $Wjj$  \\
  \hline
  $mini$-$iso > 0.9$                              &  0.87 &  0.93 &          0.0038 &         0.0023 &  0.95  \\
  \hline
  $x_\mu > 0.5$                                   &  0.87 &  0.91 &          0.0373 &         0.0353 &  0.96  \\
  \hline
  combined $x_\mu$ and $mini$-$iso$               &  0.83 &  0.89 &          0.0014 &         0.0009 &  0.93  \\
  \hline
  \DR$_{b\mu}\times \frac{p_T}{\rm TeV}$ $< 0.8$  &  0.97 &  0.97 &          0.9970 &         0.9980 &  0.45  \\
  \hline \hline
  all combined                                    &  0.81 &  0.86 &          0.0013 &         0.0009 &  0.39  \\
  \hline
\end{tabular}
\end{center}
\label{tab:1TeVdijet}
\end{table}

\begin{table}[t]
\caption{Nominal efficiencies at 2 TeV.}
\begin{center}
\begin{tabular}{ | c | c | c | c | c | c | c }
  \hline
                                                  & $t_L$ & $t_R$ &  \PYTHIA\ dijet & \HERWIG\ dijet & $Wjj$  \\
  \hline 
  $mini$-$iso > 0.9$                              &  0.86 &  0.93 &          0.0026 &         0.0024 &  0.95  \\
  \hline
  $x_\mu > 0.5$                                   &  0.84 &  0.88 &          0.0405 &         0.0445 &  0.95  \\
  \hline
  combined $x_\mu$ and $mini$-$iso$               &  0.79 &  0.86 &          0.0012 &         0.0010 &  0.92  \\
  \hline
  \DR$_{b\mu}\times \frac{p_T}{\rm TeV}$ $< 0.8$  &  0.93 &  0.96 &          0.9936 &         0.9945 &  0.27  \\
  \hline \hline
  all combined                                    &  0.73 &  0.82 &          0.0011 &         0.0009 &  0.21  \\
  \hline
\end{tabular}
\end{center}
\label{tab:2TeVdijet}
\end{table}

\section{$W$-strahlung}

\label{sec:Wstrahlung}

High-$p_T$ events with $W$-bosons constitute the second major source of background.  A $W$ produced in close proximity to a jet looks practically identical to a boosted top.  In fact, such a configuration becomes progressively more common as the momentum scale is increased well above the $W$ mass, since quarks produced at such high energies can radiate $W$s much like gluons or photons.%
\footnote{They may also radiate a $Z$.  This case should be largely dealt with using an explicit dimuon $Z$ veto.}  These $W$-strahlung emissions are similarly dominated by soft and collinear regions of phase space, but with $m_W$ acting as a physical regulator.

The extent to which this might pose a problem depends crucially on the probability of $W$ emission.  At $p_T \gg m_W$, the emission probability from an individual quark line (multiplying by the 50\% chance that it is left-handed chirality) can be estimated as
\beq
P(W{\rm\mbox{-}strahlung}) \approx \frac14 \, \frac{\alpha_2}{\pi} \, \log^2\frac{p_T}{m_W}.  \label{eq:Wrate}
\eeq
This should be multiplied by the 11\% branching fraction to muons.  At $p_T = 1$ TeV, the emission probability is about 2\%, with 0.2\% for emitting in the muon mode.  While this is a small number, the rate of hard quark production at the LHC exceeds that of top production at equivalent energies by several orders of magnitude.  Moreover, $W$-strahlung becomes increasingly important at higher $p_T$'s, in part because of the log-squared growth, but mostly since valence quark scattering turns off more slowly than any other process.

In this section, we will more carefully categorize this background, and consider how it can be ameliorated through simple cuts without significantly affecting the boosted top signal.  However, as with the heavy flavor background, we postpone discussion of absolute rates until section \ref{sec:resonances}, where we investigate the impact of $W$-strahlung on the \ttbar\ invariant mass spectrum.

With emission probabilities at the percent scale, $W$-strahlung at TeV momenta is still a highly perturbative process.  We model it at leading order with $(W^\pm\rightarrow\mu^\pm\nu)jj$ using \MadGraph/\MadEvent\ {\tt 4.4.13} (+\PYTHIA)~\cite{Alwall:2007st,Frederix:2007gi} at 14 TeV collision energy.  To force the events into the $p_T$ range of our analysis, and to avoid QCD singularities associated with soft/collinear regions for the two partons, we demand that the hardest parton is above 450 GeV, the second hardest above 50 GeV, and we place a $k_T$ cut of {\tt xqcut} $> 30$ GeV.  We place no cuts on the lepton nor the neutrino.  The events are subsequently showered, hadronized and reconstructed as in subsection \ref{subsec:reco}.  The final cross section passing our reconstruction criteria is fairly insensitive to the detailed values of our generator-level cuts.\footnote{Also, using a fully matched sample incorporating $(W^\pm\rightarrow\mu^\pm\nu)j$ shows no significant change.}  We stress that $W$-strahlung cannot be accurately modelled with the simple $2\rightarrow2$ production of $Wj$ dressed with QCD radiation, as would be obtained with \PYTHIA\ or \HERWIG\ standalone programs.  In particular, our analysis here is essentially orthogonal to that in \cite{Thaler:2008ju}.

The first tactic usually considered for discriminating a $Wj$ system from a semileptonic top is to apply an invariant mass cut of some kind.  Indeed, this has been done using transverse mass in  \cite{Baur:2007ck,Baur:2008uv}, and full neutrino reconstruction in \cite{Barger:2006hm,Agashe:2006hk}.  However, \met\ could realistically turn out to be unreliable for precision top reconstruction in this regime, and we are led to consider alternative variables which use only visible particles.  To this end, we capitalize on the angular features of $W$-strahlung.  $W$s can be emitted at arbitrary angles, with a would-be collinear singularity cut off at an angle characterized by $m_W/p_T$.  Boosted tops, on the other hand, typically decay inside of a cone characterized by the somewhat larger angle $m_t/p_T$.  At a bare minimum, then, one can veto events where the $W$ and the ``$b$-jet'' are too far apart to look like a boosted top decay.  However, at first glance, these rough estimates would still suggest that the peak of $W$-strahlung angles lies well within the distribution for top decay angles.  Fortunately, this picture is slightly misleading, and the $W$-strahlung emission actually peaks at an angle approximately five times larger than $m_W/p_T$.

This last observation is straightforward to understand.  Soft and collinear singularities originate at the diagrammatic level from intermediate propagators going almost on-shell.  The quite large singularity from the squared propagator denominator is highly (but incompletely) suppressed by polarization and phase space terms in the numerator, leaving over the usual singularities in emission angle and energy.  As soon as we move into a region of phase space where the denominator is significantly modified due to $m_W$ effects, these numerator terms act to shut off the emission rate.%
\footnote{This story does not describe longitudinal emissions, which instead turn off due to Goldstone boson equivalence in exactly those regions where transverse emissions are enhanced.  The longitudinal component is subdominant except at angles smaller than the angle where transverse emissions peak.  It is never a large contribution in an absolute sense.}  In the small-angle limit the denominator takes the form
\beq
z(1-z)\theta^2 p^2 \; + \; 2(1-z)p\left(\sqrt{z^2 p^2 + m_W^2}-z p\right) \; + \; m_W^2,  \label{eq:denom}
\eeq
where $\theta$ is the final angle between the $W$ and quark, $p$ is the original quark momentum, and $z$ is the fraction of this momentum carried by the $W$.  (This definition of $z$ extends down to zero, unlike fractional {\it energy} with a massive emission.)  Emission rates look similar to those in the massless limit only when the first term in this equation dominates.

\begin{figure}[tp]
\begin{center}
\epsfxsize=0.44\textwidth\epsfbox{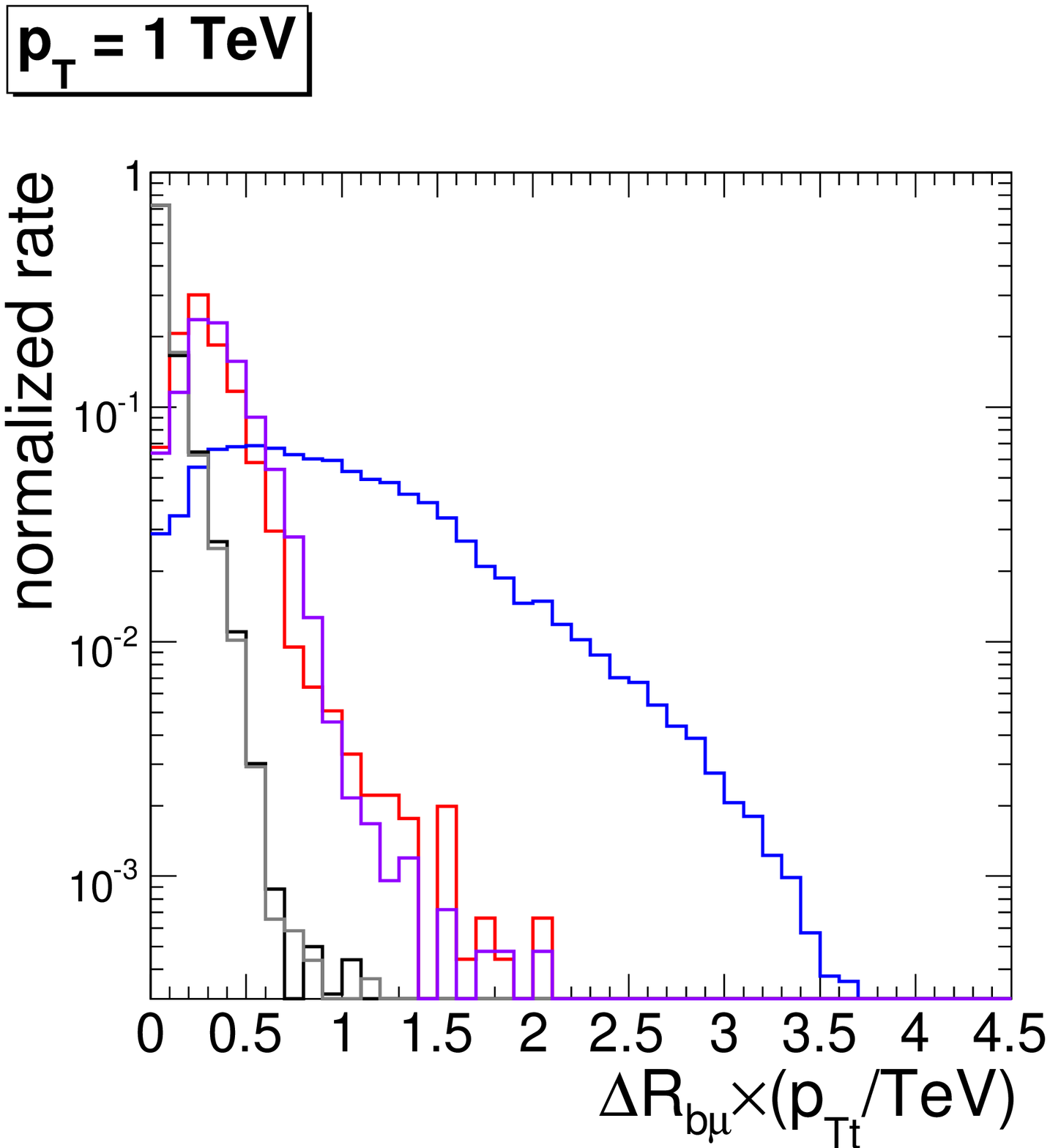}
\epsfxsize=0.44\textwidth\epsfbox{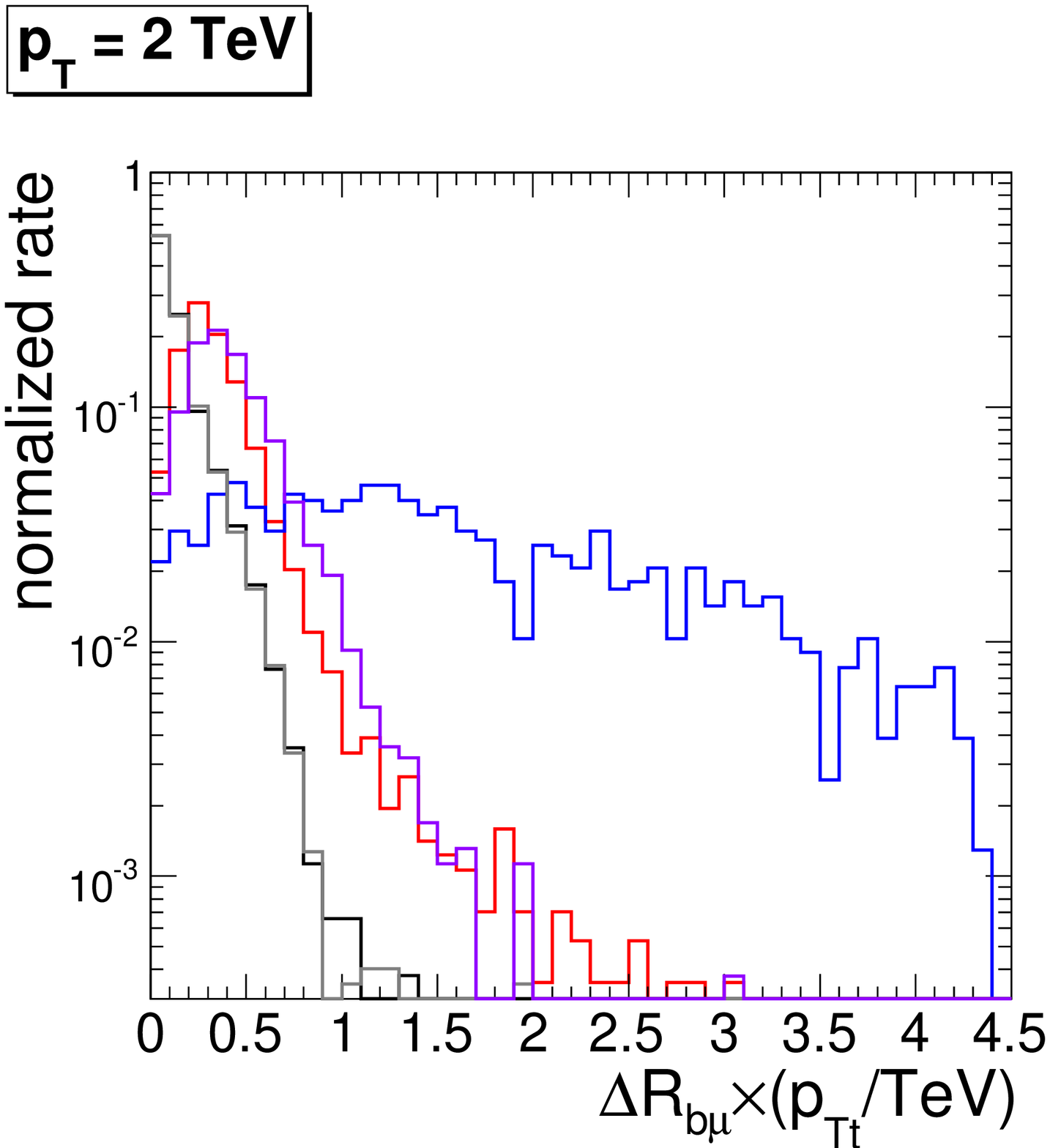}
\caption{Distributions of $\Delta R_{b\mu}\times(p_{Tt}/{\rm TeV})$ for 1 TeV and 2 TeV semileptonic top candidates reconstructed from LH chiral \ttbar\ (purple), RH chiral \ttbar\ (red), PYTHIA QCD (black), HERWIG QCD (grey), and $Wjj$ (blue).}
\label{fig:DRblScaled}
\end{center}
\end{figure}

For strictly collinear emission, the first term in Eq.~(\ref{eq:denom}) vanishes, leading to the usual dead-cone phenomenon, though now due to massive emissions instead of massive emitters.  In fact, the first term is strictly smaller than the other two for all $z$ when $\theta < 2m_W/p$.  For larger emission angles, the range of $z$ for which the first term can dominate starts to gradually open up.  Integrated over $z$, the emission rate as a function of $\theta$ ultimately peaks at an angle $O(1)$ beyond $2m_W/p$.  The predicted offset of the $W$-strahlung emission peak from the top peak is nicely illustrated in the distributions obtained from \MadGraph, displayed in Fig.~\ref{fig:DRblScaled}.  This shows the \DR\ between the muon and \bjet\ candidate at reconstructed top $p_T$ of 1 TeV and 2 TeV, with the horizontal scale multiplied by $p_T$/TeV to undo the shrinking of angles with top boost.  Note that for $p_T \gg m_W$, the muon becomes an excellent tracer of its parent $W$'s flight direction, so it inherits the features under discussion.

As a simple, experimentally-robust method of eliminating $W$-strahlung events, we therefore propose placing an {\it upper} cut on \DRbmu\ scaled with $1/p_T$.  We call this ``anti-isolation.''  Since genuine top quarks will experience shrinking angular scales as $p_T$ increases, the efficiency for tops stays roughly constant.  The probability for a quark emitting a $W$ near the dead cone region is also relatively insensitive to $p_T$.  However, the {\it total} $W$ emission rate, integrating over all angles, increases logarithmically with $p_T$ (Eq.~\ref{eq:Wrate}).  Since we define the efficiency for $W$-strahlung as the passing fraction of reconstructed $W$+jets events (appropriately binned in $p_T$), the cut naively becomes more effective at higher $p_T$.

We list the relative acceptances obtained from our simulations in Tables \ref{tab:1TeVdijet} and \ref{tab:2TeVdijet}, placing the anti-isolation cut
\beq
\DeltaR_{b\mu}\times(p_{Tt}/{\rm TeV}) < 0.8,  \label{eq:nominalWstrahlung}
\eeq
where for $p_{Tt}$ we use the $p_T$ of the recoiling hadronic top candidate.\footnote{More general classes of events could have additional sources of missing energy, the simplest example being dileptonic \ttbar.  Obtaining a measurement of the full semileptonic top $p_T$ may then be difficult or impossible.  The obvious alternative is to instead use the total {\it visible} $p_T$ of the semileptonic top candidate, namely of the muon plus the nearby jet.  We find that this provides comparable discrimination power.  However, it is important to realize that in cases where the semileptonic top $p_T$ is not fully measured, we should also be worried about discriminating against $W$-strahlung jets of {\it different} total $p_T$.  We do not perform such comparisons here.}  With this cut, we can eliminate an $O(1)$ fraction of the $W$-strahlung background while losing only a few percent of the signal.  We also show the effect of combining this cut with the heavy flavor cuts.  The two sets of cuts are fairly uncorrelated.

\begin{figure}[tp]
\begin{center}
\epsfxsize=0.44\textwidth\epsfbox{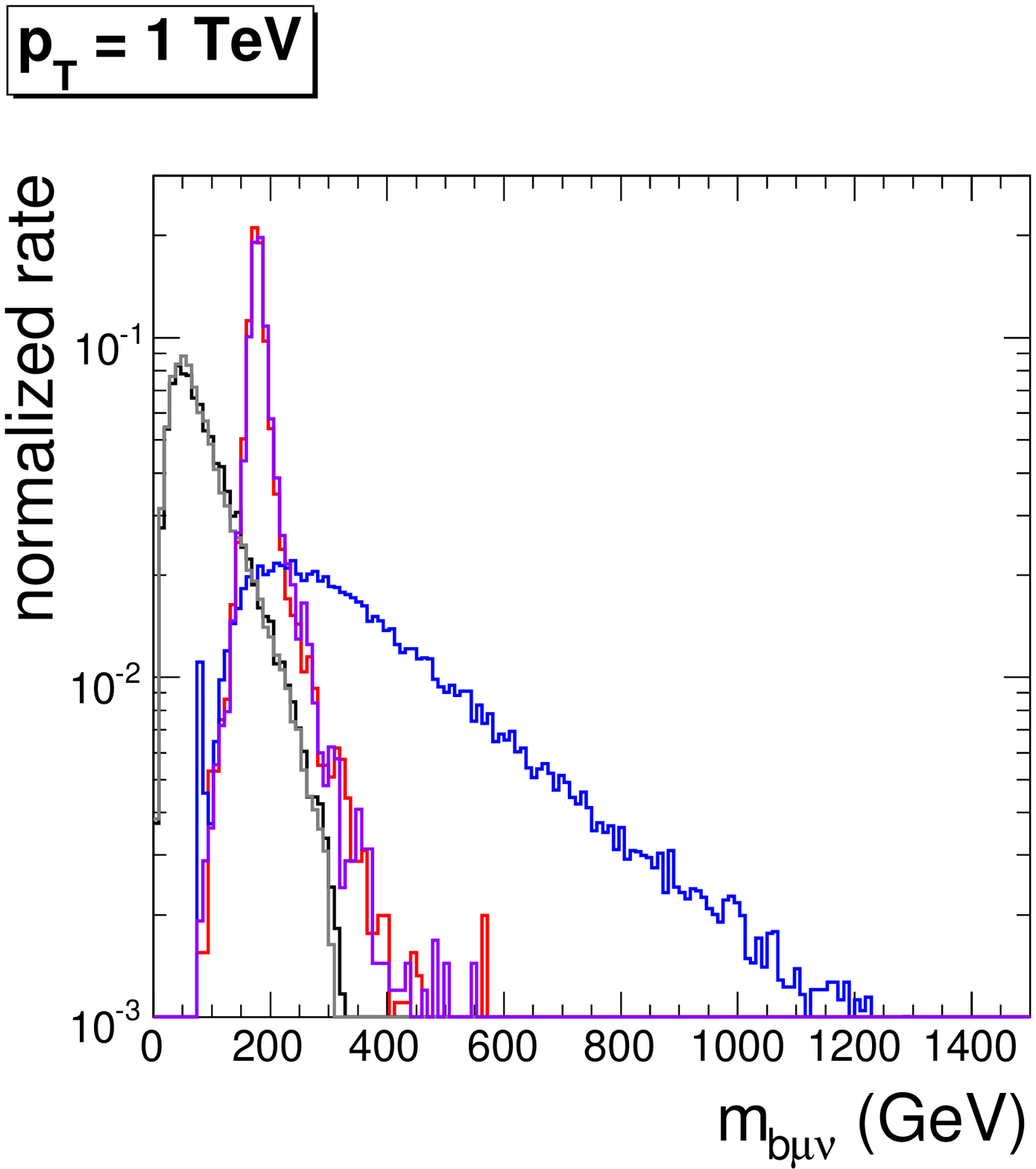}
\epsfxsize=0.44\textwidth\epsfbox{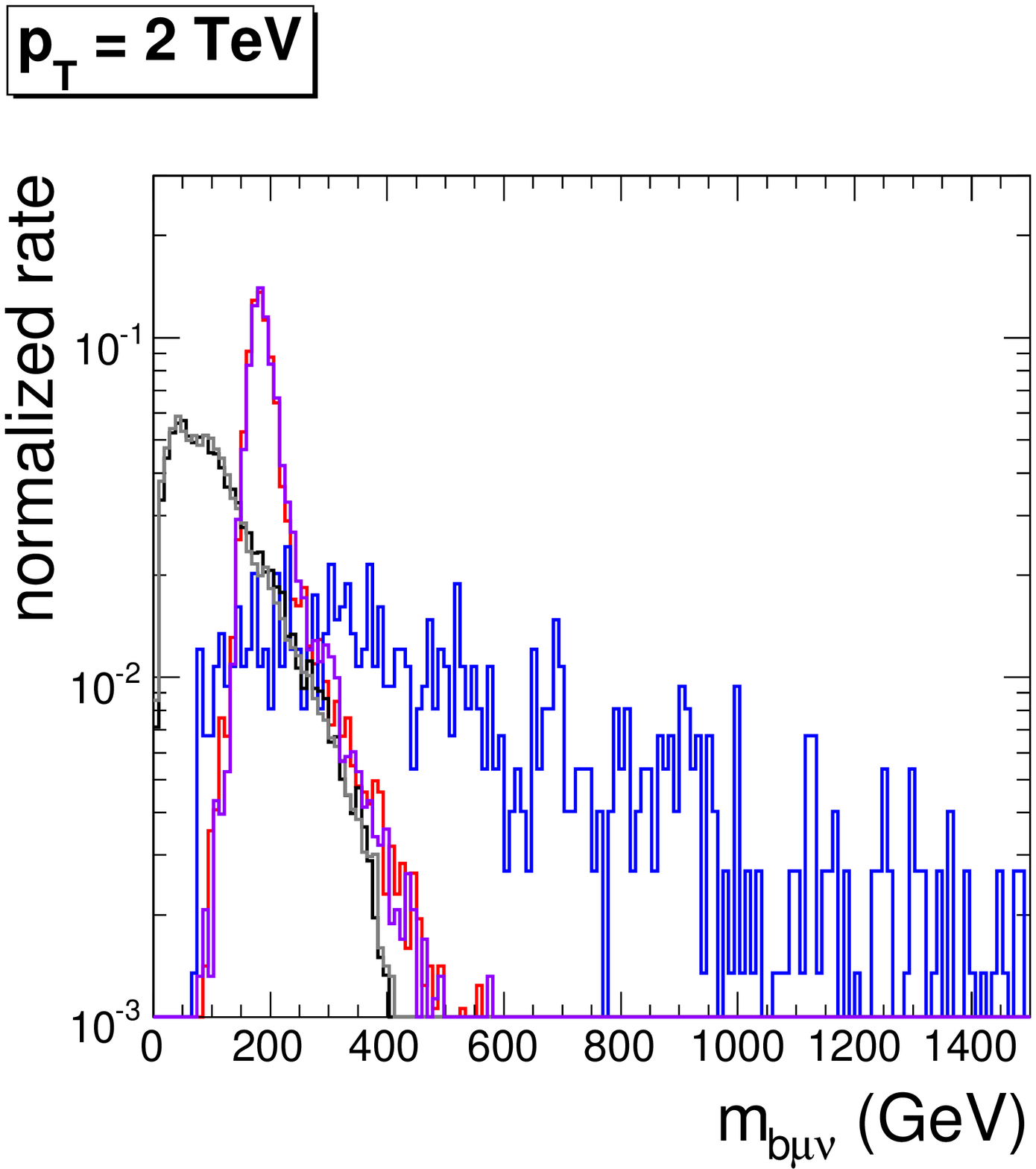}
\caption{Distributions of idealized $m_{b\mu\nu}$ (with perfectly-measured neutrino) for 1 TeV and 2 TeV semileptonic top candidates reconstructed from LH chiral \ttbar\ (purple), RH chiral \ttbar\ (red), PYTHIA QCD (black), HERWIG QCD (grey), and $Wjj$ (blue).}
\label{fig:mtopIdeal}
\end{center}
\end{figure}

\begin{figure}[tp]
\begin{center}
\epsfxsize=0.44\textwidth\epsfbox{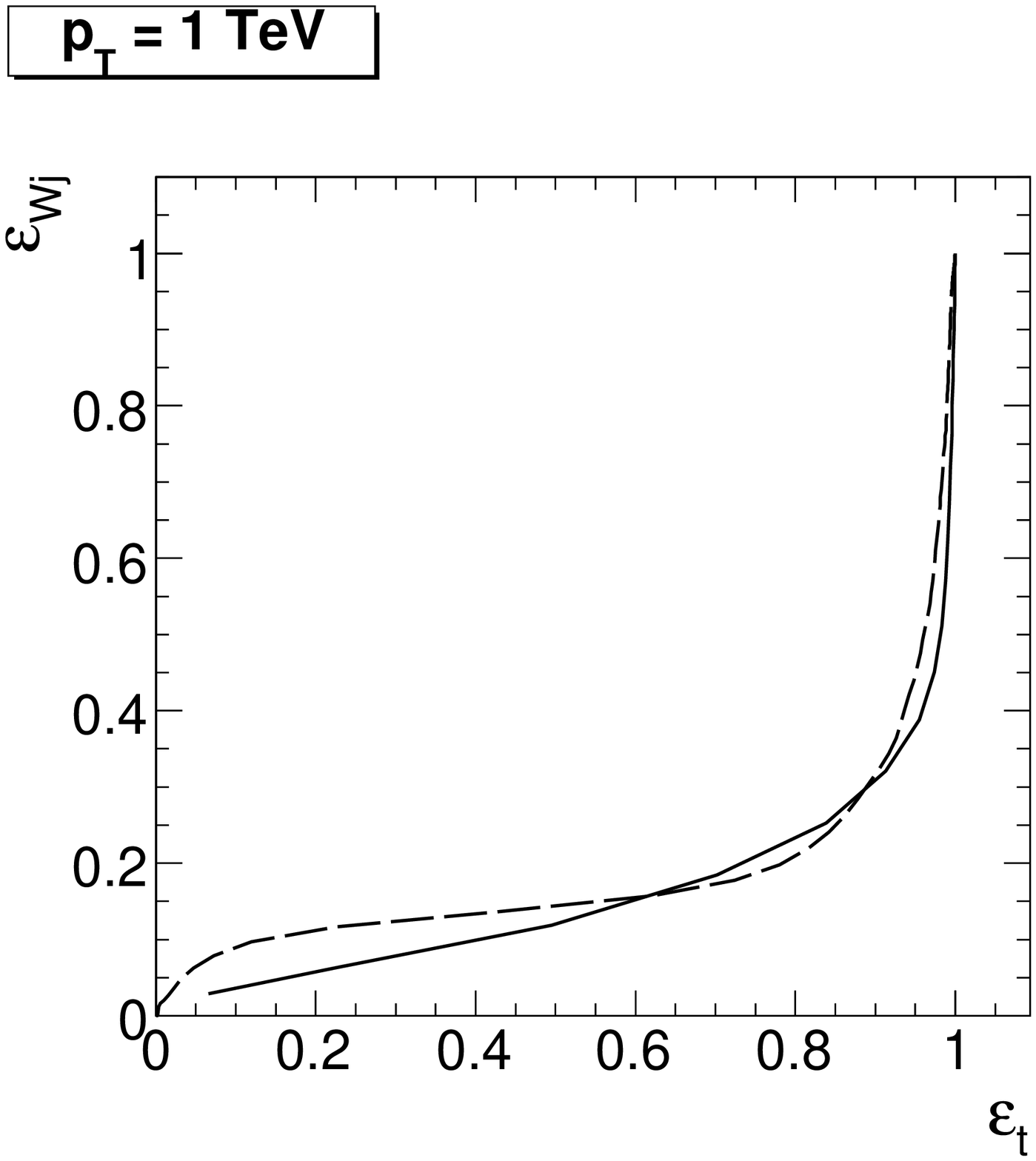}
\epsfxsize=0.44\textwidth\epsfbox{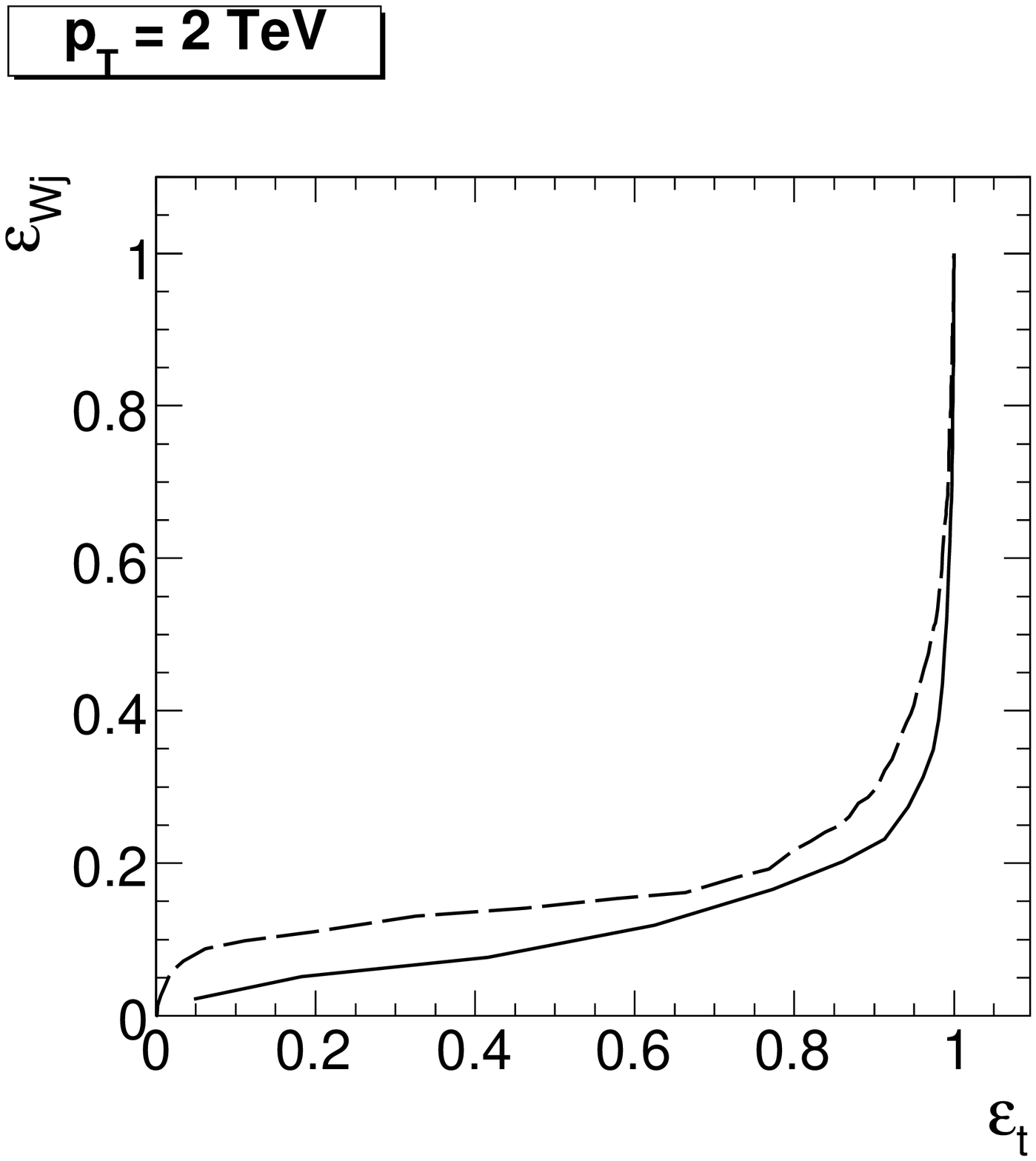}
\caption{Efficiency for jets with $W$-strahlung (reconstructed as candidate semileptonic top-jets) versus efficiency for helicity-averaged top-jets, scanning over independent one-sided cuts on $\Delta R_{b\mu}\times p_{Tt}$ (solid) and idealized $m_{b\mu\nu}$ (dashed).}

\label{fig:effWstrahlung}
\end{center}
\end{figure}

To get some idea of how anti-isolation fares against more traditional mass-based cuts, we construct a highly idealized top invariant mass variable, incorporating the exact neutrino 3-vector.  The distributions of this variable are displayed in Fig.~\ref{fig:mtopIdeal}.\footnote{A small spike at $m_W$ is visible for the $Wjj$ sample in the 1 TeV panel.  This is from rare events where our reconstruction mistakes a hard photon radiated off of the muon as the candidate \bjet.  (Statistics are not high enough at 2 TeV to see this as cleanly.)  Of course, such a feature would likely not appear with more realistic reconstruction criteria.}  No energy smearing has been applied, so the spread in the reconstructed invariant mass for real tops is entirely due to particle sampling within the \bjet.  This is some combination of out-of-range particles, uncorrelated radiation from the collision, energy lost to neutrinos in semileptonic $B$-hadron decay, and leakage of top-FSR into the \bjet\ reconstruction.  The last is the biggest effect as $p_T$ increases, but note that our jet clustering size is coarsely shrinking with $p_T$ scale, so that we automatically try to stay in the top's dead cone at some level.  In any case, our smearing of the reconstructed top mass peak is likely extremely conservative.  To compare with anti-isolation, we construct discriminator curves for \DRbmu$\times p_T$ and $m_{b\mu\nu}$ at 1 TeV and 2 TeV.  The cuts start at large values and scan down to zero.  We show the results of the scans in Fig.~\ref{fig:effWstrahlung}.  The performance is clearly comparable, and even slighly better for \DRbmu$\times p_T$ for most of the range.

As with heavy flavor, more aggressive cuts are possible, but we have attempted to understand how much can be done with minimal sensitivity to detector effects and with minimal loss of signal.  Specifically, we chose to sit near the bend of the background vs. signal efficiency curve in Fig.~\ref{fig:effWstrahlung}.  Other variables may be worth exploring, as well.  In particular, it is still possible to fold in \met\ to get a more complete picture of the kinematics.  This may afford some additional discriminating power, though our own further investigations using the complete neutrino 3-vector (after application of the $\Delta R_{b\mu}\times p_{Tt}$ cut) suggest that the potential gain may not be large.  The relative momentum of the muon in the $\mu$+$b$ system ($z_\mu$) is another simple variable~\cite{Thaler:2008ju}, but we do not find that it gives a distinctive distribution compared to real tops.  In fact, the $z_\mu$ distribution from $W$-strahlung closely mimicks that of left-chirality tops, as we will see below.

\section{Searching for Multi-TeV \ttbar\ Resonances in $\mu$+Jets}

\label{sec:resonances}

The observations above suggest that it will be possible to identify boosted semileptonic tops with high purity while maintaining high efficiency.  As an illustrative example of the effectiveness of our proposed minimal set of cuts, we present a simple search analysis for spin-1 \ttbar\ resonances in the \mujets\ channel, with the assumption that the LHC will ultimately reach its design energy of 14 TeV.  We restrict the discussion to resonances with pure left-handed or pure right-handed couplings, in order to highlight possible chirality biases in the reconstruction and to address the possibility of measuring these couplings independently.  We perform our analyses for narrow resonances such as from weakly-coupled models, as well as for 15\%-width resonances such as would arise in strongly coupled models where top is partially composite.

Event reconstruction follows the same logic as in subsection \ref{subsec:reco}.  For events passing the reconstruction, we further demand that the semileptonic top candidate satisfies our nominal heavy flavor and $W$-strahlung cuts, as in Eqs.~\ref{eq:nominalhf} and~\ref{eq:nominalWstrahlung}, respectively.

Until this point, we have been treating all energy measurements as exact, and have been explicitly avoiding the use of missing energy for reconstruction of the full neutrino momentum.  Here, we incorporate smearing of particle energies and a simplistic neutrino reconstruction, in order to roughly model their effects on the reconstructed resonance peaks.

We smear jet energies according to the CMS physics TDR~\cite{Bayatian:2006zz}:
\beq
\frac{\sigma(E)}{E} = \frac{5.6\;{\rm GeV}}{E} \oplus \frac{1.25\;{\rm GeV}^{1/2}}{\sqrt{E}} \oplus 0.033.
\eeq
Technically, this formula only applies to iterative cone jets with $R = 0.5$, but we do not anticipate any major change when going over to C/A jets of comparable size.  Using the CMS parametrization is the conservative choice, since the ATLAS hadronic calorimeter is expected to have better resolution~\cite{:1999fq}.  The first term also incorporates fluctuations from particle sampling within the jet area, which are already implicit in our jet construction.

To smear the muon energy, we use the parametrization from the CMS muon TDR~\cite{Bayatian:1997ki}, increasing the coefficient by 25\% to better match the resolution curves presented in the more recent physics TDR~\cite{Bayatian:2006zz}:
\beq
\frac{\sigma(E)}{E} = 0.05 \; \sqrt{ \frac{E}{{\rm TeV}} }.
\eeq
This assumes that global muon reconstruction is possible within TeV-scale top-jets. In principle, stiffer muon tracks should be easier to trace from the outer detectors into the inner detectors, and, as we have emphasized, muons from top decay will be mini-isolated at tracker level.

\begin{figure}[tp]
\begin{center}
\epsfxsize=0.44\textwidth\epsfbox{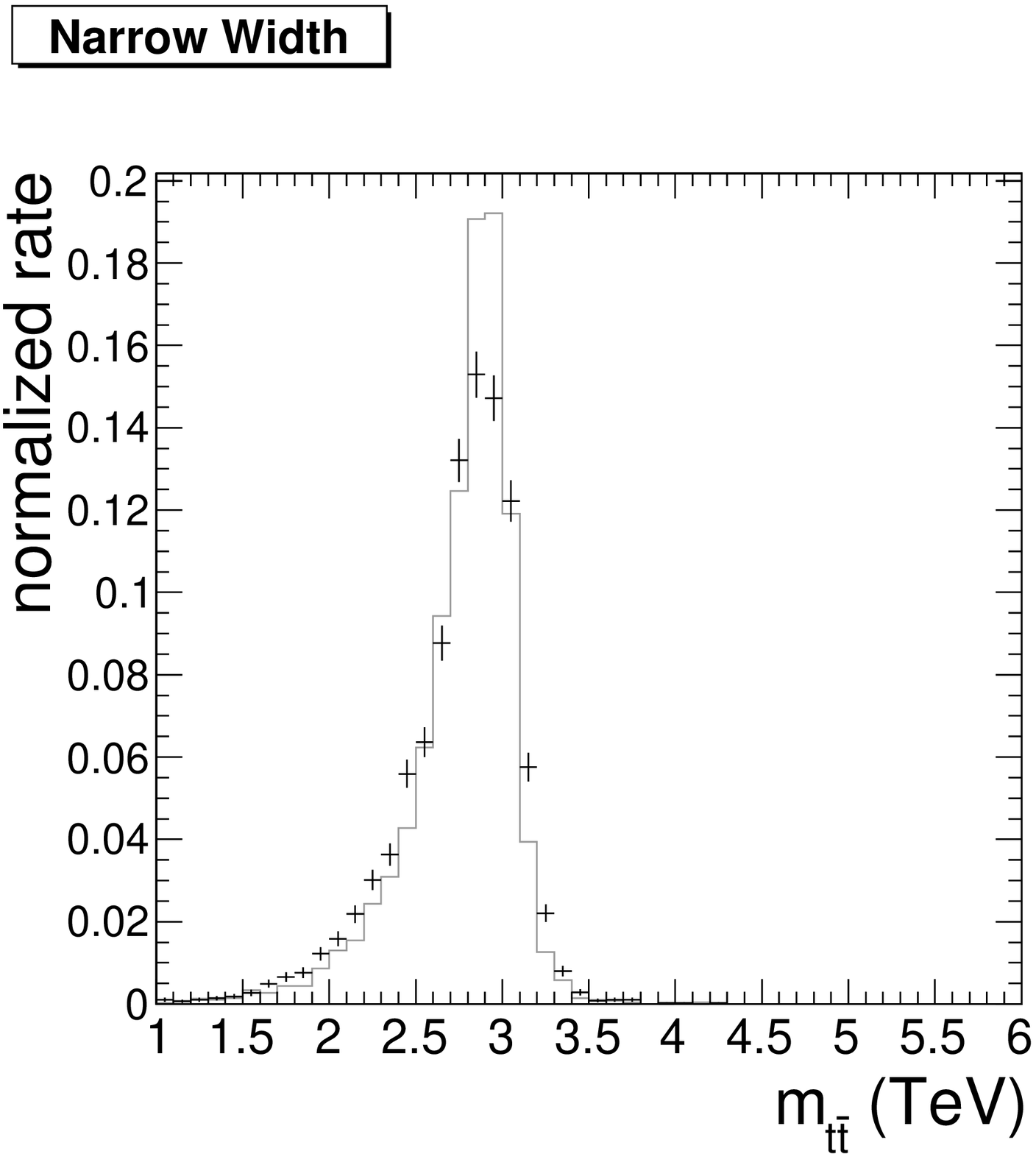}
\epsfxsize=0.44\textwidth\epsfbox{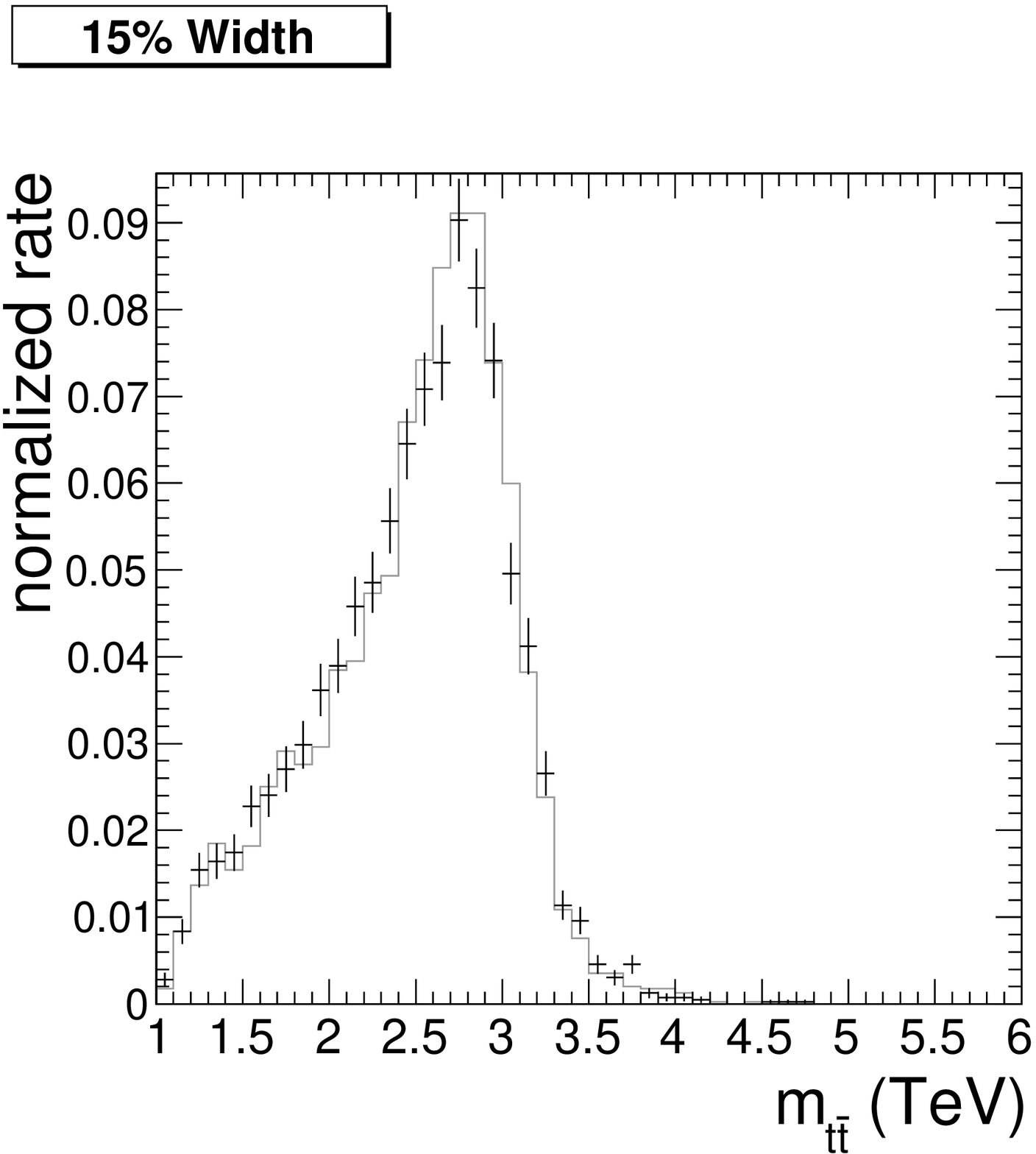}
\caption{Reconstructed \ttbar\ invariant mass spectra of narrow (left) and 15\%-width (right) spin-1 resonances with $M = 3$ TeV, coupled to $t_R$.  We ignore any interference effects with continuum production.  The black histogram represents our nominal treatment of missing energy, with error bars from monte carlo statistics.  The continuous grey histogram (with error bars suppressed) represents a more idealized reconstruction using the exact neutrino 3-vector.}
\label{fig:resonance}
\end{center}
\end{figure}

We define \met\ to balance the leading reconstructed objects in the event after energy smearing.  We include the hadronic top-jet, the \bjet, the muon, and the leading remaining jet, if there are any additional jets found.\footnote{This jet would usually come from hard ISR or FSR, and incorporating it reduces the occurence of outliers in the mass spectrum due to badly mismodelled \met.}  We reconstruct the $p_z$ of the neutrino by merely assigning it the same $\eta$ as the muon, rather than attempting to impose a $W$ mass constraint.  The \ttbar\ system is then simply the sum of the reconstructed hadronic and semileptonic tops.%
\footnote{A more sophisticated analysis would also identify possible FSR from the tops before they decay.  For example, one could simply incorporate the leading jet within some reasonable \DR\ from either of the reconstructed tops.  Given the level of our estimated smearing, we do not find that this procedure offers significant improvements in our final resonance lineshapes.  However, incorporation of top-FSR would be very useful to study in a more realistic analysis.}  In Fig.~\ref{fig:resonance} we show the effects of our reconstruction and smearings on narrow and 15\%-width spin-1 resonances with $M = 3$ TeV.  The plot includes our nominal procedure, as well as one incorporating an exactly measured neutrino 3-vector, for comparison.  The instrumental width is clearly dominated by our jet and muon energy smearing, and not the missing energy reconstruction.

\begin{figure}[tp]
\begin{center}
\epsfxsize=0.44\textwidth\epsfbox{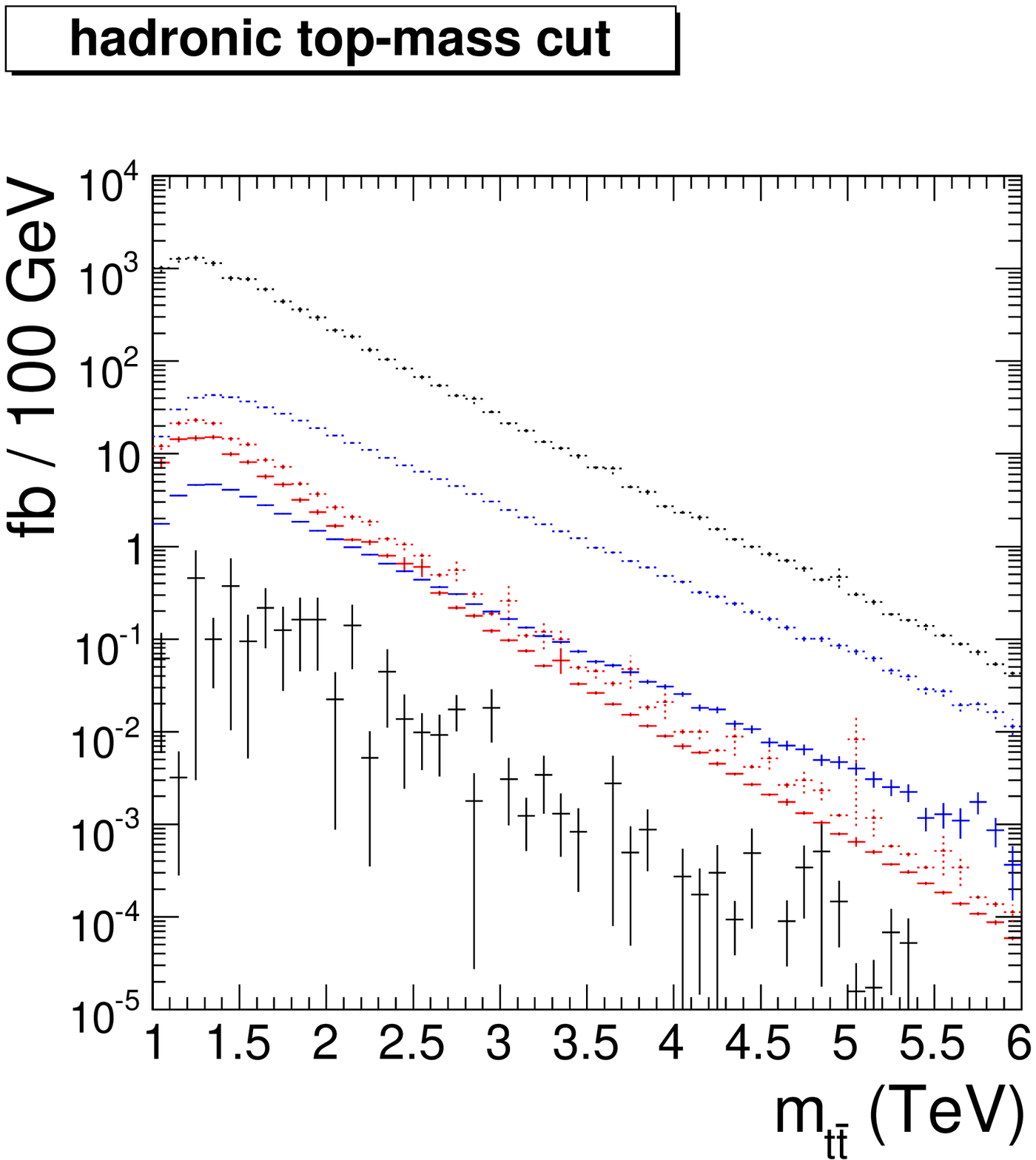}
\epsfxsize=0.44\textwidth\epsfbox{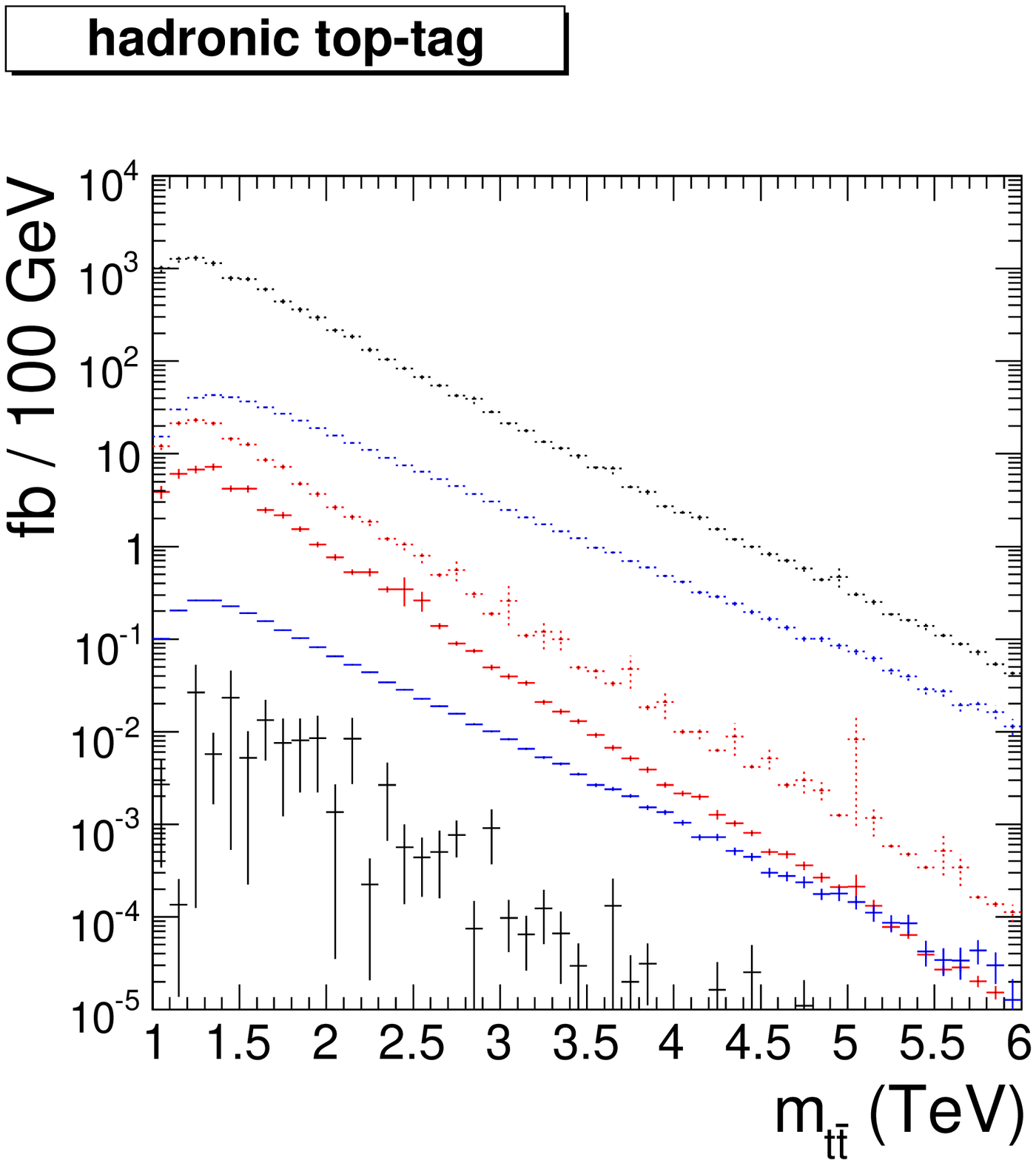}
\caption{Backgrounds in the \ttbar\ invariant mass spectrum at leading order:  PYTHIA dijet (black), $Wjj$ (blue), and continuum \ttbar\ in the $\mu$+jets channel (red).  Reconstruction of the event invariant mass is as described in the text.  The left panel displays events in which the hadronic top candidate passes a simple top mass cut (also described in the text).  The right panel displays events in which the hadronic top candidate passes the top-tag of~\cite{Kaplan:2008ie}.  In each panel, the dashed histograms display the background spectra before application of any cuts on the hadronic or leptonic top candidates beyond the basic reconstruction criteria of subsection~\ref{subsec:reco}.  Error bars reflect monte carlo statistics.}
\label{fig:backgrounds}
\end{center}
\end{figure}

Further purification of the signal can be achieved by applying cuts on the hadronic top-jet.  The hadronic calorimeter top-tag of~\cite{Kaplan:2008ie} is an obvious option, provided we are willing to tolerate a factor of $\gsim 2$ loss of signal.  As an alternative, we also consider a looser ``top-tag'' in the form of a simple top-mass cut.  A hadronic top-jet candidate passes the cut if its (unsmeared) invariant mass is between 120 GeV and 300 GeV.  This cut passes $(80 \sim 90$)\% of top-jets, $(20 \sim 30)$\% of light quark jets, and $(30 \sim 50)$\% of gluon jets, depending on $p_T$.  Figure~\ref{fig:backgrounds} shows the leading-order backgrounds in the \ttbar\ invariant mass spectrum using our full reconstruction and cuts, including both methods of hadronic top-tagging.\footnote{In \cite{Kaplan:2008ie}, we and our collaborators presented top-tagging efficiencies up to $p_T = 2$ TeV.  For the current study, we also consider top-jets at even higher $p_T$.  For example, at $p_T = 3$ TeV, the efficiencies for tops, light quarks, and gluons are about $2$\%, $0.2$\%, and $0.5$\%, respectively.  The efficiency for finding tops decreases dramatically due to its decay products falling into adjacent or identical calorimeter cells.  We expect that more sophisticated techniques, incorporating information from the electromagnetic calorimeter and tracker, could likely improve the situation for tops with multi-TeV transverse momenta.}  In order to maintain good statistics on the dijet and $W$-strahlung backgrounds, we utilize weighted tags on quark and gluon jets (binned in $p_T$), based on studies of independent \PYTHIA\ dijet samples.

For our nominal analysis, we use the hadronic top-mass cut instead of the full top-tag, in order to keep good signal efficiency across the invariant mass spectrum.  The price we pay for this high efficiency is that the reducible $W$-strahlung background begins to dominate above about $2.5$ TeV invariant mass.  However, signal reach across the entire invariant mass spectrum is still better than the full top-tag analysis.  Obviously, a more efficient top-tag or high-$p_T$ $b$-tag could offer further improvements.

\begin{figure}[tp]
\begin{center}
\epsfxsize=0.44\textwidth\epsfbox{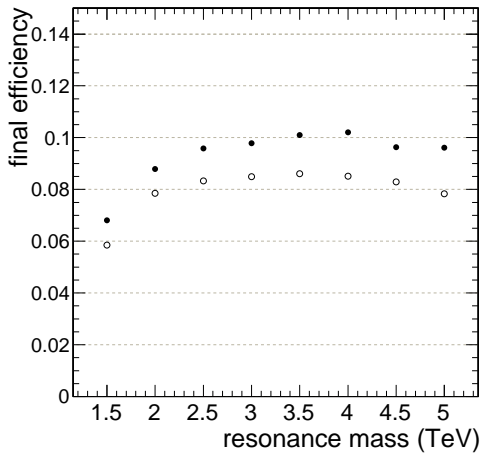}
\caption{Nominal final efficiencies for narrow spin-1 resonances decaying into \ttbar.  (The $\mu$+jets branching fraction sets an upper bound of about 15\%.)  Open circles represent resonances with purely left-handed couplings, and solid circles represent resonances with purely right-handed couplings.}
\label{fig:efficiencies}
\end{center}
\end{figure}

We can already get some sense for the efficiency of our cuts to pass genuine top pairs by looking at the irreducible \ttbar\ background in the left panel of Fig.~\ref{fig:backgrounds}.  Over half of the reconstructed $\mu$+jets events survive the cuts.  We also plot, in Fig.~\ref{fig:efficiencies}, the final efficiencies of narrow chiral spin-1 resonances decaying to \ttbar, normalized with respect to the total $\sigma\times$BR(\ttbar).  Perfect efficiency for our analysis would correspond to the $\mu$+jets branching ratio, or approximately 15\%.  Our procedures achieve efficiencies of close to 10\%.\footnote{The slight difference between the efficiencies for reconstructing left-chirality and right-chirality resonances mostly owes to the semileptonic top reconstruction efficiency, as in Tables~\ref{tab:1TeVdijet} and~\ref{tab:2TeVdijet}.  The hadronic top-mass cut (as well as the full hadronic top-tag) is relatively insensitive to top polarization.}

\begin{figure}[tp]
\begin{center}
\epsfxsize=0.44\textwidth\epsfbox{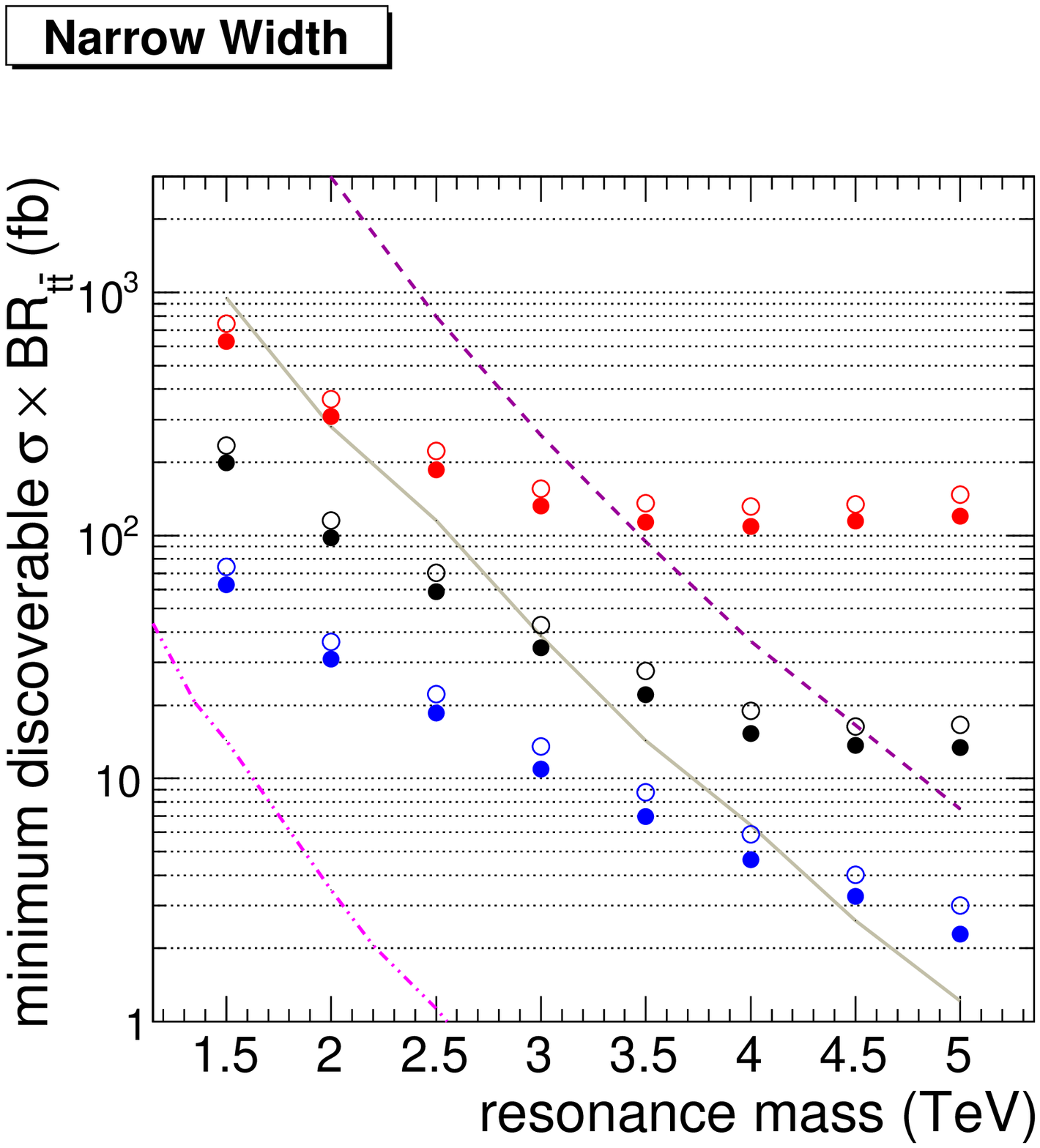}
\epsfxsize=0.44\textwidth\epsfbox{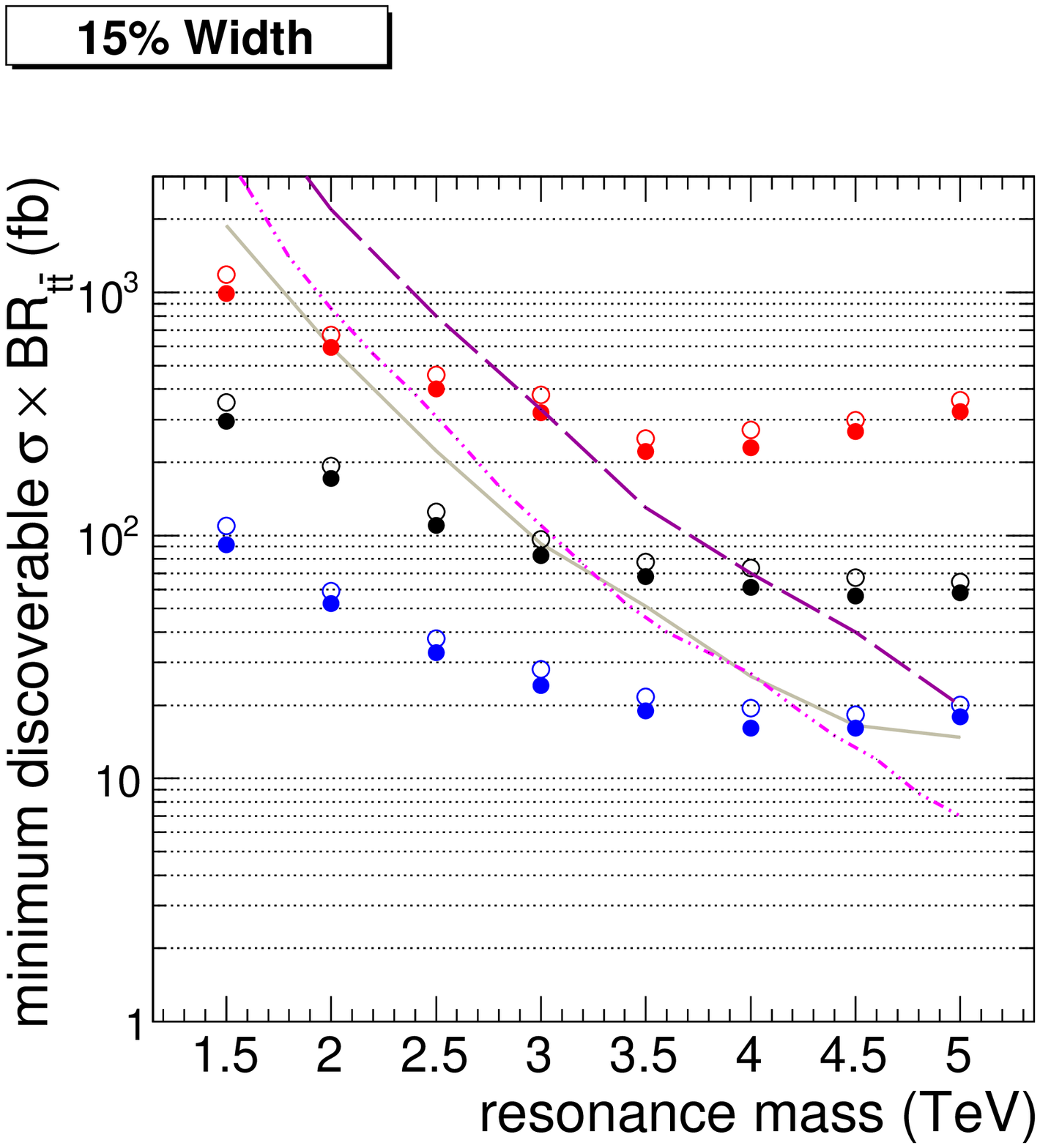}
\caption{The minimum discoverable $\sigma\times$BR$(t\bar{t})$ for narrow (left) and 15\%-width (right) multi-TeV spin-1 resonances.  Open circles are pure left-handed chiral resonances, and solid circles are pure right-handed.  Luminosities displayed are 1 fb$^{-1}$ (red), 10 fb$^{-1}$ (black), and 100 fb$^{-1}$ (blue).  The grey line represents the $\sigma\times$BR$(t\bar{t})$ at which $N_S/N_B = 1$ within the signal windows (for the right-chirality resonances only).  Also displayed are several models.  In the narrow resonance plot (left):  a pure $B-L$ gauge boson with $g_{B-L} = 0.2$~\cite{Basso:2010pe} (pink dot-dashed), and a KK gluon of the Little Randall Sundrum model, with $y = 5$~\cite{Davoudiasl:2008hx,Davoudiasl:2009jk} (purple dashed).  In the 15\%-width resonance plot (right):  estimates for the KK gluon of~\cite{Agashe:2003zs} taken from~\cite{Agashe:2006hk} (pink dot-dashed) and~\cite{Lillie:2007yh} (purple dashed).}
\label{fig:reach}
\end{center}
\end{figure}

Given the high reconstruction efficiency and manageable level of reducible background, we can achieve promising sensitivity to new resonances.  We display the estimated reach for $\sigma\times$BR(\ttbar) in Fig.~\ref{fig:reach}.  To perform the estimate, we simply count events within coarsely optimized mass windows constructed about each resonance.  For the narrow resonances, we take $m_{t\bar t}$ within $\pm10\%$ of the physical resonance mass, and for the 15\%-width resonances we take $\pm15\%$.\footnote{These windows are actually somewhat asymmetric about the reconstructed resonance peaks, which are a few percent below the original mass.  Since the background spectrum is a falling function, the windows are therefore slightly biased towards the regions with higher signal-to-background ratio.}  A given production cross section is considered discoverable if the expected number of signal plus background events is at least 5$\sigma$ above the background-only prediction according to Poisson statistics (asymptotically, $N_S/\sqrt{N_B} > 5$) and $N_S > 10$.  This estimate essentially represents the most optimistic signal reach, assuming well-controlled systematic errors.  To indicate the possible relevance of background uncertainties, we also plot the $\sigma\times$BR(\ttbar) at which $N_S/N_B = 1$ within the signal windows.  Superimposed on the plots are several model predictions.

We highlight our conclusions for the lightest KK gluon of~\cite{Agashe:2003zs}, a particle which can also be viewed as an excitation of a strongly-coupled sector which generates a composite Higgs boson.  Assuming the model curves of~\cite{Agashe:2006hk} or~\cite{Lillie:2007yh}, respectively, we find that a 100 fb$^{-1}$ run of 14 TeV LHC can potentially discover a KK gluon up to about 4.5 TeV or 5.0 TeV.  In previous studies, the discovery reach for this particle had been estimated to be less than 4 TeV~\cite{Agashe:2006hk,Baur:2008uv} at 100 fb$^{-1}$.

\begin{figure}[tp]
\begin{center}
\epsfxsize=0.44\textwidth\epsfbox{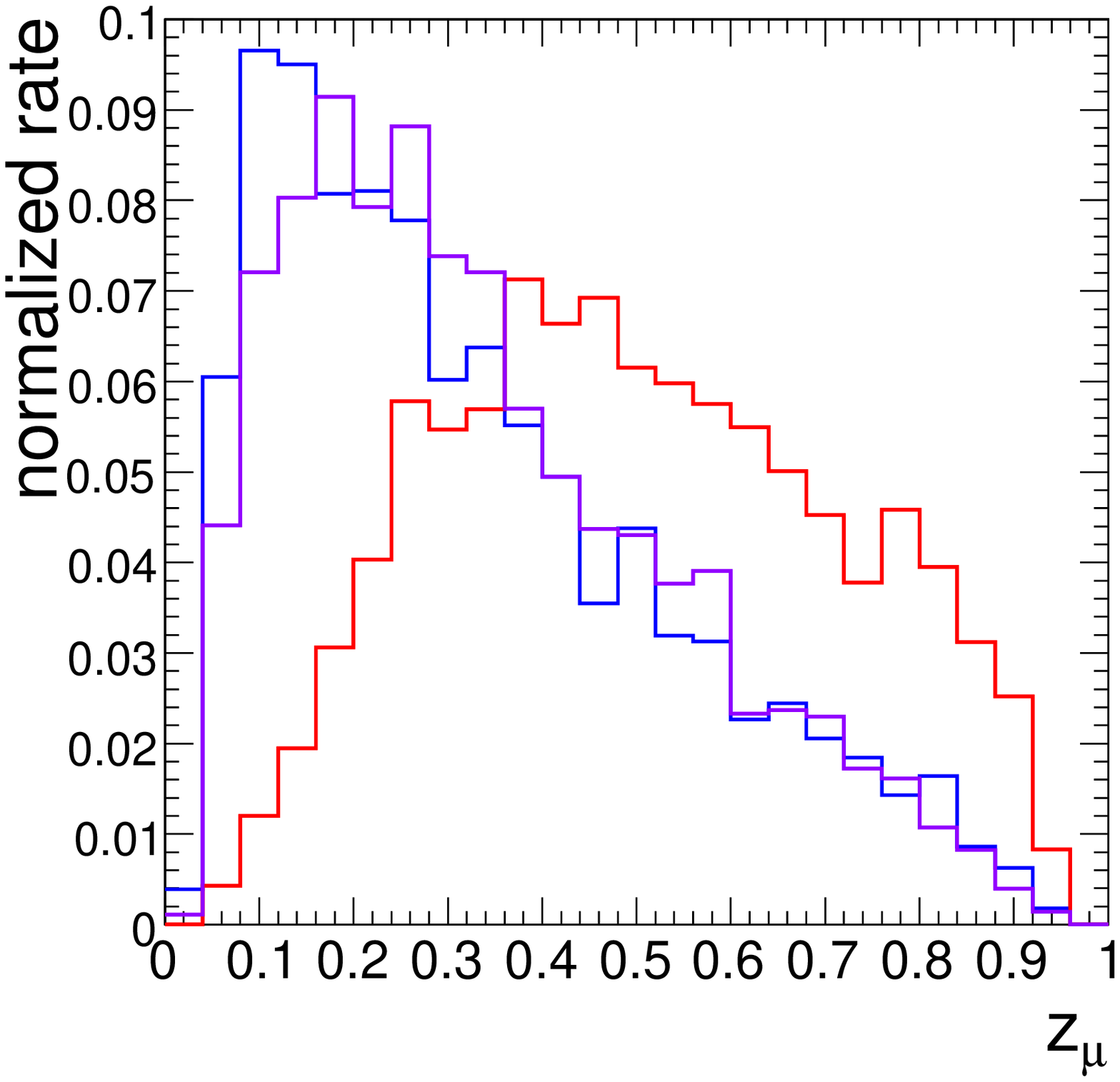}
\caption{Normalized distributions of the polarization-sensitive variable $z_\mu$ for the narrow 3 TeV left-chirality resonance (purple), narrow 3 TeV right-chirality resonance (red), and $Wjj$ background (blue), all reconstructed within the \ttbar\ invariant mass window $m_{t\bar t} = [2700,3300]$ GeV.}
\label{fig:polarization}
\end{center}
\end{figure}

As pointed out in~\cite{Agashe:2006hk,Lillie:2007yh}, heavy composite resonances may have enhanced couplings to $t_R$ versus to $t_L$, and this bias may be experimentally observable in semileptonic decays.  A general analysis of simple boosted top ``polarimeter'' variables was performed in~\cite{Shelton:2008nq}, and polarization effects were further explored in the context of boosted hadronic tops in~\cite{Krohn:2009wm}.  Here, we do not attempt to quantify the quality of polarization measurements for different models.  However, to illustrate the extent to which polarization information is preserved by our procedures, we display in Fig.~\ref{fig:polarization} the distribution of the polarization-sensitive variable $z_\mu$ in the mass window $m_{t\bar t} = [2700,3300]$ GeV.  The samples compared include the narrow 3 TeV left- and right-chirality resonances, as well as the $W$-strahlung background.  The $z_\mu$ distribution from the continuum \ttbar\ background is close to the average between the two chiral resonances, and we have omitted it for clarity.  It is worth noting that if we had chosen to use $z_\mu$ (or an analogous quantity) as a discriminator variable for eliminating backgrounds, this polarization-sensitive structure would have been degraded.

We can make two immediate observations regarding polarimetry measurements.  First, left- and right-chirality resonances are clearly distinguishable from each other, and nearly reproduce the distributions predicted in~\cite{Shelton:2008nq}.\footnote{The predicted $t_R$ shape in~\cite{Shelton:2008nq} is actually slowly monotonically rising as $z_\mu$ approaches one.  However, our reconstruction methods in subsection~\ref{subsec:reco} demand the presence of a hard $b$-jet candidate to pair with muon.  This shuts off the rate near $z_\mu = 1$.  A more permissive reconstruction would likely yield a more distinctive $z_\mu$ distribution for $t_R$.}  If the signal-to-background ratio is favorably large, and statistics are high, discriminating between the two cases should be straightforward.  However, we can also observe that the final $z_\mu$ distribution of $W$-strahlung events closely mimicks that of $t_L$.  Since $W$-strahlung is the main background for resonances above about $2.5$ TeV in our nominal analysis, care would need to be taken in making reliable polarization measurements.  Of course, the $W$-strahlung contamination could be significantly reduced by applying stricter cuts, such as a full hadronic top-tag on the recoiling top candidate (Fig.~\ref{fig:backgrounds}), though at an additional cost of signal efficiency.

\section{Conclusions and Outlook}

\label{sec:conclusions}

Efficient identification of boosted top quarks will be important for various well-motivated new physics searches at the LHC.  In this paper, we have explored techniques to optimize boosted top identification for the cleanest case of semileptonic decays in the muon mode, focusing on observables that should be relatively robust against detector effects.  In particular, we have avoided the use of missing energy and $b$-tags, in contrast to more standard search strategies.  Nonetheless, we have found excellent background rejection while keeping very high signal efficiency, as evidenced by the rates in Tables~\ref{tab:1TeVdijet} and~\ref{tab:2TeVdijet}.

The most subtle and ubiquitous background to boosted semileptonic tops is ordinary QCD jets with hard embedded muons, dominantly from heavy flavor decays.  For TeV-scale jets, the heavy flavor usually originates from gluon splittings in the parton shower, as opposed to prompt production in the hard interaction.  We have compared several candidate discriminator variables, and found that the most powerful is a tracker-level ``mini-isolation'' using a small cone which shrinks with increasing muon $p_T$.  This is in contrast to more traditional lepton isolation, which tallies tracker and calorimeter energy within a cone comparable to a fixed jet clustering radius, e.g., $R=0.4$.  Combining with an additional cut on the muonic mass-drop variable $x_\mu$ of~\cite{Thaler:2008ju} (or the $\Delta R$ between the muon and the center of its associated jet), signal-to-background can potentially be purified by almost 1000:1, with only $O(10\%)$ loss of signal.  This attenuation of background is in addition to the initial requirement that the QCD jet contains a hard muon, which occurs with percent-scale probability.  For resonance searches in the \ttbar\ invariant mass spectrum, this level of QCD jet rejection reduces the dijet contribution to essentially negligible level.  The dijet background will remain negligible or modest if even a small fraction of our claimed discrimination power can be achieved at the LHC.

Jets containing heavy flavor produce muons through off-shell $W$-boson emission.  However, it is also possible for a light quark to directly emit an on-shell $W$-boson, i.e.\ to undergo $W$-strahlung.  These emissions are collinear-enhanced for TeV-scale jets, occurring with percent-scale probability.  In the absence of $b$-tagging, a light quark jet with $W$-strahlung can look very similar to a top-jet.  Nonetheless, top-jets are more kinematically constrained than $W$-strahlung, and we have investigated a very simple cut that appears to capture most of the available kinematic discriminating power:  an ``anti-isolation'' cut on the {\it maximum} $\DeltaR$ between the muon and its accompanying jet.  This can achieve $O(1)$ purification of signal-to-background with a few percent loss of signal.

As a case study, we have investigated the performance of a minimal set of cuts in the context of a multi-TeV \ttbar\ resonance search at a 14 TeV LHC.  The major remaining backgrounds are irreducible \ttbar\ continuum and reducible $W$-strahlung, with the latter becoming dominant above about $2.5$ TeV.  The final resonance reconstruction efficiencies, {\it including} the branching fraction to $\mu$+jets, are close to 10\%, suggesting much better sensitivity than has been previously estimated~\cite{Agashe:2006hk,Baur:2008uv}.  In particular, we find that a warped KK gluon could be discovered at masses above 4 TeV with 100 fb$^{-1}$ of data.  Left-chirality and right-chirality resonances can be reconstructed with similar efficiency, and our cuts preserve much of the polarization-sensitive kinematics.  However, the remaining $W$-strahlung background mimicks left-chirality tops, possibly complicating polarization measurement for high mass resonances unless further cuts are applied.

Our study leaves open several important issues.  The most pressing is the possible role of detector effects.  The simple anti-isolation cut for $W$-strahlung is likely the easiest to implement without a detailed understanding of the detector.  However, construction of the mini-isolation variable will require tracking to work quite reliably within the core of a jet, which could be a rather crowded environment.  We suspect that this possible tracking breakdown would not prove to be an insurmountable problem.  Genuine top-jets are characterized by a muon slightly offset from the bulk of the jet activity, and should be much easier to track and mini-isolate.  Our main worry then becomes whether we can obtain a reliable measure of the activity around the muons within QCD jets.  But even if detailed momentum measurements become difficult in this case, the nearby density of tracking hits can likely serve as an effective supplementary discriminator.  Also, we have not modelled possible tracker signals originating from neutral pions, due to photons showering in the tracker material.  This additional activity will likely make muons from QCD even easier to discriminate.  Ultimately, determination of the true effectiveness of mini-isolation or analogous discriminators will require more detailed detector simulations and analysis of high-energy LHC data.  Our particle-level results suggest that such pursuits are highly motivated.

We can also consider various avenues for improvement of boosted semileptonic top identification, in particular the incorporation of $b$-tags, the use of \met\ for detailed kinematic reconstruction, and the possibility of using the electron decay mode.

We have neglected $b$-tagging since it may be difficult to implement with good efficiency at high $p_T$, and in any case it cannot be reliably modelled at the level of our analysis.  However, even a loose displaced vertex tag could be useful for improving discrimination against $W$-strahlung jets, which are mainly associated with light quarks.  For example, an additional $O(1)$ reduction of the $W$-strahlung background would be quite beneficial for searches in the \ttbar\ invariant mass spectrum, assuming that the signal efficiency can be kept high.  

Missing energy, on the other hand, may be of limited utility.  We have managed to essentially eliminate heavy flavor backgrounds without ever referencing \met.  For $W$-strahlung events, after passing our anti-isolation cut the kinematics are already extremely similar to boosted tops.  Given the susceptibility of \met\ to fluctuations in visible energy measurements, it will be surprising if significantly better kinematic discrimination can be achieved.

We have focused entirely on the case of semileptonic top decays into muons, but we might also hope to recover a signal in decays into electrons.  Work at ATLAS suggests that electron reconstruction within top-jets should be possible with $O(1)$ efficiency~\cite{Brooijmans:2009boo}, with well-controlled QCD backgrounds.  We may therefore hope to ultimately achieve somewhat greater signal reach beyond a muon-only analysis by incorporating these decays.  They could be especially relevant in classes of events where we search for more than one boosted top decaying semileptonically.

Finally, for searches involving both boosted semileptonic and hadronic tops, we note that hadronic top-tagging can also have a significant impact.  In particular, it serves as an additional way to eliminate $W$-strahlung background from the \ttbar\ invariant mass spectrum in the $\mu$+jets channel, by top-tagging the recoiling jet.  We already saw the potential for the hadronic calorimeter top-tag of~\cite{Kaplan:2008ie} in the right-hand panel of Fig.~\ref{fig:backgrounds}, albeit at high cost to the signal for large invariant masses.  More sophisticated top-tagging techniques are clearly worth investigating.  Also, beyond the $\mu$+jets search, hadronic top-tagging opens the possibility of performing an all-hadronic search for \ttbar\ resonances~\cite{Kaplan:2008ie,CMS:2009b}.  While $\mu$+jets is still the optimal resonance discovery mode, owing to its modest contamination from QCD background, reproducing a claimed discovery in the all-hadronic channel would serve as a powerful cross-check.


\acknowledgments{We have benefited from useful conversations with David E.~Kaplan, Muge Karagoz, Petar Maksimovic, Kirill Melnikov, Salvatore Rappoccio, and Morris Swartz.  K.R. is supported in part by National Science Foundation grant NSF-PHY-0401513.  B.T. is supported by the Leon Madansky Fellowship and by Johns Hopkins University grant \#80020033.}

\appendix


\bibliography{lit}
\bibliographystyle{apsper}

\end{document}